\documentclass[iop]{emulateapj}
\usepackage[colorlinks]{hyperref}
\hypersetup{colorlinks=true,linkcolor=red,citecolor=blue,filecolor=blue,urlcolor=blue,}
\usepackage{amsmath}
\usepackage{algorithm}
\usepackage{amsfonts}
\usepackage{mathrsfs}
\usepackage{bm}
\usepackage{rotating}
\usepackage{color}
\usepackage{graphicx}
\usepackage{subfigure}
\usepackage{lineno}
\usepackage{savesym}
\savesymbol{tablenum}
\restoresymbol{SIX}{tablenum}
\def\cha{\textit{Chandra}}
\def\XMM{{XMM-\textit{Newton}}}
\def\NuSTAR{\textit{NuSTAR}}
\def\Nu{\textit{NuSTAR,}}
\def\bat{{{\it Swift}-BAT}}
\def \XSPEC {{\tt XSPEC}}

\def \MYTorus {{\tt MYTorus}}
\def \borus {{\tt borus02}}
\def \bntorus {{\tt BNtorus}}
\def \ngc {NGC 3081}
\def \eso {ESO 565-G019}

\begin{document}
\title{COMPTON-THICK AGN IN THE \NuSTAR\ ERA VII. A JOINT \Nu\ \cha\ and \XMM\ ANALYSIS OF TWO NEARBY, HEAVILY OBSCURED SOURCES}

\author{A. Traina\altaffilmark{1,2}, S. Marchesi\altaffilmark{2,3}, C. Vignali\altaffilmark{1,2}, N. Torres-Albà\altaffilmark{3},  M. Ajello\altaffilmark{3}, A. Pizzetti\altaffilmark{3}, R. Silver\altaffilmark{3}, X. Zhao\altaffilmark{3}, T. Ananna\altaffilmark{4,5}, M. Baloković\altaffilmark{5,4}, P. Boorman\altaffilmark{8,9}, P. Gandhi\altaffilmark{8}, R. Gilli\altaffilmark{2}, G. Lanzuisi\altaffilmark{2}}

\altaffiltext{1}{Dipartimento di Fisica e Astronomia (DIFA), Università di Bologna, via Gobetti 93/2, I-40129 Bologna, Italy}
\altaffiltext{2}{Istituto Nazionale di Astrofisica (INAF)-Osservatorio di Astrofisica e Scienza dello Spazio (OAS), via Gobetti 101, I-40129 Bologna, Italy}
\altaffiltext{3}{Department of Physics and Astronomy, Clemson University, Kinard Lab of Physics, Clemson, SC 29634, USA}
\altaffiltext{4}{Department of Physics, Yale University, P.O. Box 201820, New Haven, CT 06520-8120, USA} 
\altaffiltext{5}{Yale Center for Astronomy and Astrophysics, P.O. Box 208121, New Haven, CT 06520, USA}
\altaffiltext{6}{Harvard \& Smithsonian, 60 Garden Street, Cambridge, MA 02138, USA}
\altaffiltext{7}{Black Hole Initiative at Harvard University, 20 Garden Street, Cambridge, MA 02138, USA}
\altaffiltext{8}{Department of Physics and Astronomy, University of Southampton, Southampton SO17 1BJ, UK}
\altaffiltext{9}{Astronomical Institute, Academy of Sciences, Boční II 1401, CZ-14131 Prague, Czech Republic}

\begin{abstract}
We present the joint \textit{Chandra}, XMM-\textit{Newton} and \NuSTAR\ analysis of two nearby Seyfert galaxies, \ngc\ and ESO 565-G019. These are the only two having \cha\ data in a larger sample of ten low redshift ($z \le 0.05$), candidates Compton-thick Active Galactic Nuclei (AGN) selected in the 15--150\,keV band with \bat\ that were still lacking \NuSTAR\ data. Our spectral analysis, performed using physically-motivated models, provides an estimate of both the line-of-sight (l.o.s.) and average (N$_{H,S}$) column densities of the two torii. \ngc\ has a Compton-thin l.o.s. column density N$_{H,z}$=[0.58-0.62] $\times 10^{24}$cm$^{-2}$, but the N$_{H,S}$, beyond the Compton-thick threshold (N$_{H,S}$=[1.41-1.78] $\times 10^{24}$cm$^{-2}$), suggests a “patchy” scenario for the distribution of the circumnuclear matter. \eso\ has both Compton-thick l.o.s. and N$_{H,S}$ column densities (N$_{H,z}>$2.31 $\times 10^{24}$cm$^{-2}$ and N$_{H,S} >$2.57 $\times 10^{24}$cm$^{-2}$, respectively). The use of physically-motivated models, coupled with the broad energy range covered by the data (0.6--70 keV and 0.6--40 keV, for \ngc\ and ESO 565-G019, respectively) allows us to constrain the covering factor of the obscuring material, which is C$_{TOR}$=[0.63-0.82] for NGC 3081, and C$_{TOR}$=[0.39-0.65] for ESO 565-G019.
\end{abstract}


%
%
\section{Introduction}\label{sec:intro}


One of the main goals of extragalactic astrophysics is to achieve a thorough knowledge of the processes responsible for the observed emission from  Active Galactic Nuclei (AGN).
Models of AGN unification \citep[e.g.,][]{antonucci1993unified, urry1995unified} require the presence of an obscuring structure (often associated with the obscuring torus) surrounding the central supermassive black hole (SMBH). Depending on the angle between the torus axis and the line-of-sight (l.o.s.) of the observer, the AGN emission will be attenuated if it intercepts the obscuring material. The AGN classification divides these sources into two main types: Type 1 and Type 2, according to the extinction, the width of the emission lines observed in their optical spectra and the shape of the continuum \citep[see, e.g.,][]{padovani2017active}. 
\par In the X-ray band, Type 1 and Type 2 AGN are generally referred to as unobscured and obscured, respectively \citep[][]{osterbrock1978observational}. The second type includes the so-called Compton-thin (N$_H$ $\sim 10^{22-24}$ cm$^{-2}$) and Compton-thick (CT, N$_H$ $\ge 10^{24}$ cm$^{-2}$) sources \citep[][]{comastri2004compton}; in this last case, the obscuring material strongly attenuates the nuclear emission below 10 keV. Studies on the AGN population have suggested that their emission can account for most of the Cosmic X-ray Background \citep[CXB, i.e., the diffuse emission observed between $\sim$0.5--500 keV,][]{gilli2007synthesis}; specifically, Type 2 AGN play an important role in shaping the CXB, as well as in the context of the AGN-galaxy co-evolution \citep[][]{treister2010major}, especially at high redshift. On the one hand, unobscured AGN contribution to the CXB is nowadays almost completely resolved into point-like sources. On the other hand, the detection of obscured AGN, which are responsible for a significant fraction of the CXB emission \citep[$\sim$40\% at the peak,][]{gilli2007synthesis, ananna19synthesis}, is challenging.

Thus, the study of CT-AGN, can provide a better characterization of  the CXB, especially around the peak \citep[$E\sim 30$ keV,][]{ajello2008cosmic}. From observations, the CT-AGN fraction at $z\sim0$ results to be $\sim$10-20\% \citep[see e.g.,][]{comastri2004compton, burlon2011three, ricci2015compton}, that is lower than the one expected from CXB population synthesis models \citep[20-50\%,][]{gilli2007synthesis, ueda2014toward, buchner2015obscuration, ananna19synthesis, Zhao2020clumpy}.
\par In order to fill the gap between observations and model predictions, a census of obscured AGN (in particular, CT-AGN) is needed, combining data at different wavelengths. In particular, since X-rays are energetic enough to penetrate the obscuring material (i.e., the torus) up to considerable amounts of column density, X-ray observations offer a unique possibility for the characterization of the inner regions of the AGN.
\par Since the effect of the absorption by the obscuring material varies with the photon energy, the radiation with energy below 10 keV becomes much more attenuated with respect to higher energy photons. For this reason, the NASA and ESA flagship X-ray telescopes, \cha\ and \XMM\ (active in the 0.3--10 keV energy band), cannot entirely characterize the spectral properties of such obscured sources at $z\sim0$. Telescopes which cover a higher energy band, such as \textit{Swift} Burst Alert Telescope \citep[BAT,][]{Barthelmy2005BAT} or the Nuclear Spectroscopic Telescope Array \citep[\textit{NuSTAR},][]{Harrison2013Nustar}, are thus required to create a more unbiased census of black holes.
\par Recent works \citep[e.g.,][]{burlon2011three, ricci2015compton} have been carried out using the \bat\ telescope data at high energies ($\sim 15-150$ keV), combined, if available, with $0.3-10$ keV data. However, newer studies \citep[e.g.,][]{marchesi2018compton} reveal, in the comparison between \bat\ and \NuSTAR\ spectra of heavily obscured AGN, the presence of an offset in the values of the photon index ($\Gamma$) and the intrinsic absorption (N$_H$), which are often overestimated when \NuSTAR\ data are not used. 
\par On the basis of these results, it is clear that a combination of high quality \cha\ or \XMM\ $0.3-10$~keV data with deep \NuSTAR\ observations in the $3-79$~keV band is needed to have a broad-band characterization of the X-ray spectrum of heavily obscured sources. Such a multi-observatory synergy provides an optimal spectral coverage for the determination of the main spectral parameters, which would not be possible without one of the two bands. In order to obtain a proper characterization of the main spectral and physical properties of obscured AGN, physically motivated models \citep[e.g., \MYTorus\ and \texttt{borus02}; ][]{murphy2009x, balokovic2018new} which make use of Monte-Carlo codes should be used to reproduce the evolution of the radiative transfer through the obscuring material. These models allow one to describe the geometrical distribution of the torus and its physical properties such as the l.o.s. and intrinsic column density (N$_{H,z}$ and N$_{H,s}$), and the torus covering factor (C$_{TOR}$).
\par In order to reach a complete census of Compton-thick AGN (in the local Universe) in the X-ray band, via the detailed study of their obscuring structure, an approved \textit{NuSTAR} project (PI: S. Marchesi; proposal number 5197) “The Compton thick AGN Legacy project. A complete set of \textit{NuSTAR}-observed nearby CT-AGNs” is being carried out. This project has multiple goals, and aims to achieve a complete X-ray characterization of the sample of 57 low-redshift CT-AGN candidates, selected from the 100-month {\it Swift}/BAT catalog, through almost simultaneous \XMM\ and \textit{NuSTAR} observations. This would allow us to obtain indications on the physical and geometrical properties of the source nuclear regions. In particular, the combined use of \XMM\ and \NuSTAR\ allows to precisely constrain physical and geometrical parameters of the obscuring torus (e.g., C$_{TOR}$), which allows us to study relations such as $L_X-C_{TOR}$, through which, and coupled with variability information, it may be possible to place constraints on the nature and geometry of the obscuring torus. Finally, another goal of this large program is the determination of the intrinsic fraction of CT-AGN, as well as the space density of these type of sources.  In past works \citep[see][]{Marchesi2017ApJ...836..116M, marchesi2018compton, Marchesi2019Nustar5, Zhao2019Nustar4, Zhao2019Nustar2} the majority of the candidate CT-AGN in the 100-month BAT sample sources have been analyzed by our group; in 2019, the last 10 sources of the sample, which were still lacking \NuSTAR\ data, have been observed using \textit{NuSTAR} and \XMM\ (when lacking). In this work we present the spectral analysis of two nearby ($z=0.008$ and $z=0.016$) candidate CT-AGN (\ngc\ and ESO 565-G019) selected from the 100-months \bat\ catalog. These are the only two objects, out of the 10 in the \NuSTAR\ program, that also have \cha\ data. For this reason, we decide to study them in a separate paper, with the goal of using \cha\ subarcsecond resolution to investigate the properties of the diffuse emission around the accreting supermassive black hole. Also, the contribution of the \cha\ data to the whole spectral counts (see Table \ref{tab:extraction}), enables us to derive the parameters of interest with higher accuracy, disentangling, for example the effect of the thermal emission from the continuum scattered by thin material, at low energies.

The rest of the sample will be analyzed in a companion paper (Torres-Albà et al. submitted). 
The paper is organized as follows: in Section \ref{sec:sample} we report the process of data reduction for the three data-sets available and the extraction of the spectra; in Section \ref{sec:spectral} we describe the different models used in the spectral analysis; in Section \ref{analysis} we report the spectral analysis with the different models and in Section \ref{discussion} we summarize our results, focusing particularly on the properties of the obscuring material. All reported uncertainties on spectral parameters are at 90\% confidence level, if not otherwise stated. The standard cosmological constants adopted are: $H_0$ = 70 km s$^{-1}$ Mpc$^{-1}$, $\Omega_{M}=0.29$ and $\Omega_{\Lambda}$ = 0.71. 
%
%
\section{Sample and data reduction}\label{sec:sample}
\ngc\ and \eso\ have simultaneous observations with \XMM\ and \NuSTAR\ (PI: Marchesi) which ensure a broad-band coverage ($\sim 0.3-70$ keV) and no variability effects. Moreover, for \ngc\ and ESO 565-G019, \cha\ archival data are also available (PIs: Maksym and Koss, respectively), which allow us, thanks to {\it Chandra}'s sub-arcsecond resolution, to detect diffuse emission from the region near the nucleus, which may be due to hot gas thermally emitting, scattering or photoionization effects. Furthermore, the availability of \cha\ data contributes to improve the counts statistics in the 0.5--7 keV band, leading to a better spectral coverage. 
\par According to the \textit{NED} morphological and spectral classification, \ngc\ is a SAB0 spiral galaxy and it is classified as a Sy 1 \citep[][]{cetty2006class}; however, \cite{esparza2018circumnuclear} claim that it has a Type 2 nucleus. \cite{ricci2017bat}, using combined XMM-\textit{Swift/XRT, Chandra} and \textit{Swift}-BAT 0.3--150 keV data, found it to be heavily obscured, having $log(N_H$/ cm$^{-2})$ $= 23.91 \pm 0.04$.
\par ESO 565-G019 is a Sy2, E type galaxy \citep[][]{devaucouleurs1991class}.
From \textit{Swift}-BAT and \textit{Suzaku} X-ray observation of \eso\, its emission results to be reflection dominated, with a column density larger than the Compton-thick threshold \citep[][]{gandhi2013reflection}. In Table \ref{tab:available_data} we report the main information on the sources' observations analyzed in this work. 

\begin{table*}[]
\renewcommand{\arraystretch}{1.5}
\center
\caption{ObsID, exposure time and start date for the \textit{Chandra}, \XMM\ and \NuSTAR\ observations of NGC 3081 and ESO 565-G019. In the exposure time column, for the \XMM\ observation the exposure time after the cleaning from background flares is reported.}
\begin{tabular}{ccccccc}
\hline
\hline
                      &  & \textbf{Instrument} & \textbf{ObsID} & \textbf{Exp. time [ks]} & \textbf{Start date}\\ \hline
\textbf{NGC 3081}     &  & \cha\            & 20622                   & 29.4                       & 2018-01-24 \\
                      &  & \XMM\         & 0852180701              & 30.0                       & 2019-12-24 \\
                      &  & \NuSTAR\             & 60561044002             & 55.6                       & 2019-12-23 \\ \hline
\textbf{ESO 565-G019} &  & \cha\            & 22248                   & 10.0                        & 2019-06-06 \\
                      &  & \XMM\         & 0852180601              & 27.0                       & 2019-12-18 \\
                      &  & \NuSTAR\             & 60561043002             & 50.4                       & 2019-12-17 \\ \hline
\end{tabular}
\label{tab:available_data}
\end{table*}

\subsection{\cha\ data reduction}\label{sec:cha}
\begin{figure*} 
\centering
\begin{minipage}[b]{.39\textwidth}
\centering
\includegraphics[width=0.979\textwidth]{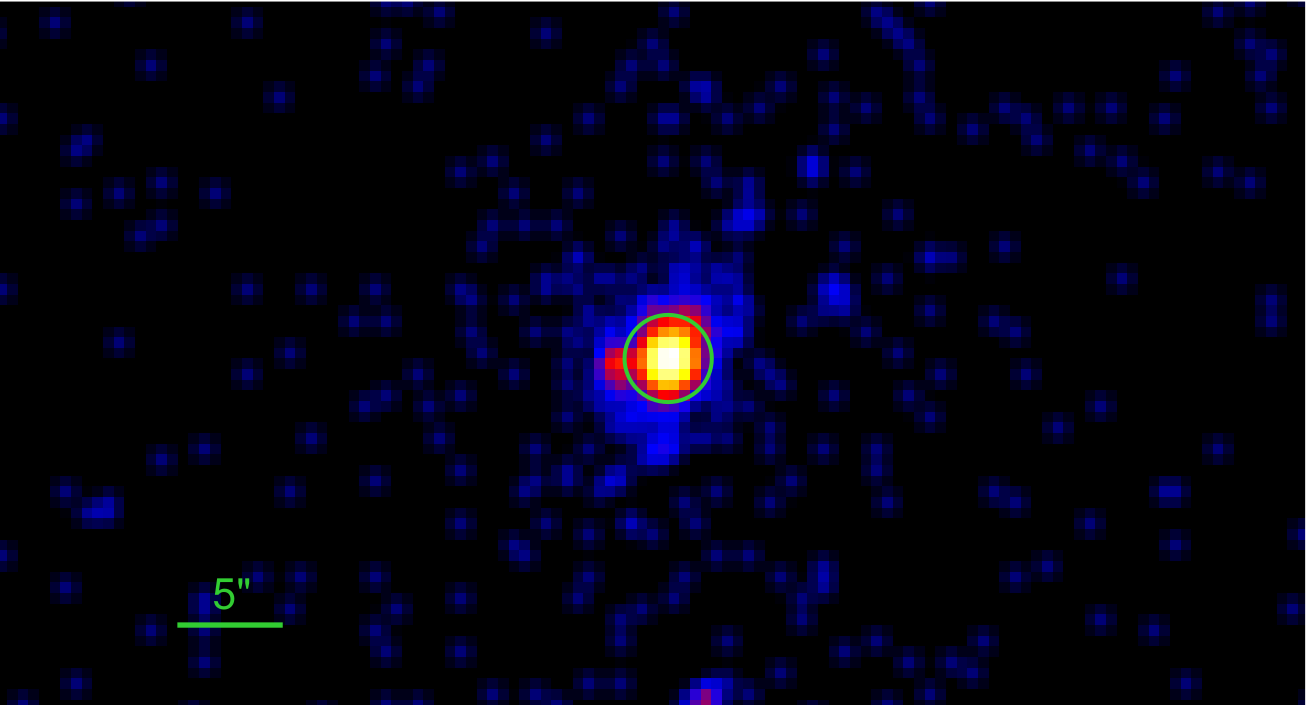}
\end{minipage}
\begin{minipage}[b]{.4\textwidth}
\centering
\includegraphics[width=1\textwidth]{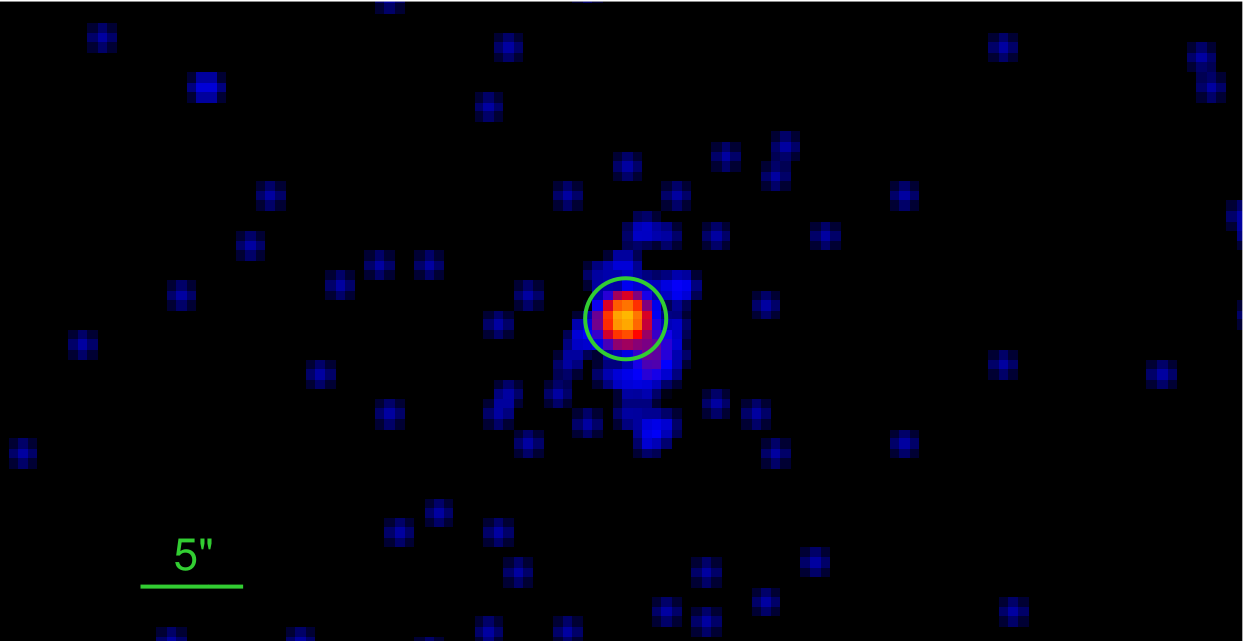}
\end{minipage}
\caption{0.3--7 keV \cha\ images of NGC 3081 (left panel) and ESO 565-G019 (right panel). In both cases the source extraction regions are reported as a circle with a radius of 2". The boxes have dimensions  43"x 26". 1" corresponds to $\sim 0.17$ kpc and $\sim 0.34$ kpc for \ngc\ and ESO 565-G019, respectively.}
\label{fig:chandra_images}
\end{figure*}

We use \cha\ archival observations with an exposure time of $\sim$ 29 ks for \ngc\ and $\sim$ 10 ks for ESO 565-G019.
The images of NGC 3081 and ESO 565-G019 in the 0.3--7.0 keV energy range are shown in Figure \ref{fig:chandra_images}. The sources in the \cha\ images are not point-like, given how the emission is extended beyond the Encircled Energy Fraction (EEF, i.e., the circular region containing a certain fraction of the counts) radius. This is due to the excellent angular resolution of the \cha\ telescope, that allows to distinguish the nuclear emission from the extended one. 
The spectra extraction has been done using the \texttt{specextract} task. This task requires the selection of an extracting region for the sources and for the background (the background region has to be unaffected by the presence of other sources). The source extraction has been chosen on the basis of the EEF for a point like source, at a fixed energy (in the case of the \textit{Chandra} HRMA a circle with a radius of about 2 arcsec contains $90\%$ of the total energy at 5 keV), so to minimize the contamination from non-nuclear emission we have chosen the energy centroid of the \cha\ image in the $E=2-7$ keV to better define the AGN.
The source regions have been chosen with a radius of 2" and the background regions have a radius of $\sim25"$.
\par Finally we bin the spectra with the \texttt{grppha} task to have at least 15 counts per bin to apply $\chi^2$ statistics. Because of the small number of spectral counts, in the case of ESO 565-G019, we have used C-stat instead of $\chi^2$ statistics in this analysis (i.e., the spectral analysis has been carried out in the Poisson regime).

\subsection{\XMM\ data reduction}\label{sec:xmm}
The \XMM\ observations, which are quasi simultaneous to the \NuSTAR\ ones (see Table \ref{tab:available_data}), have an exposure time of $\sim$ 30 ks and $\sim$ 27 ks, for \ngc\ and ESO 565-G019, respectively.
\par We create the lightcurve at energies $E>10$ keV to select a threshold to remove part of observation particularly affected by noise (i.e., particle background noise). In the case of pn, we select a value of $0.4$, and $0.35$ counts/s for MOS1 and MOS2 for \eso\ and, in the case of MOS2, we choose $0.25$ counts/s in order to remove a bright background flare for NGC 3081. In both cases the lightcurve were extracted from the whole field. We use extraction regions corresponding to an aperture which contains the $90\%$ of total energy at 5.0 keV: $40"$ in the case of MOS1, $45"$ for MOS2 and $35"$ in the pn case. For both sources, the background regions have sizes of 70, 90 and 60 arcseconds, for the MOS1, MOS2 and pn, respectively.
\par Finally, we grouped the spectra to have at least 20 counts per bin, in order to apply the $\chi^2$ statistics.

\subsection{\NuSTAR\ data reduction}\label{sec:nustar}
The \NuSTAR\ observations have exposures of $\sim$ 56 ks and $\sim$ 50 ks, for \ngc\ and ESO 565-G019, respectively.
The first step of \NuSTAR\ data reduction is the creation of the calibrated events files which will be cleaned and used to produce an exposure map. This process can be carried out using \texttt{nupipeline}.
\par The choice of the source extraction regions was made selecting four circles with different radii and inspecting the background counts and signal-to-noise ratio (S/N) variations. In the first case (NGC 3081), the data show a linear increase of the S/N with the region diameter (almost until 80", with increasing background contribution), so the choice fell on the Half Power Diameter (HPD) which contains $50\%$ of the encircled energy fraction and corresponds to a radius of 60". In the second case (ESO 565-G019), exceeding 40" leads to an increase of the background contribution. Lastly, we extract the spectra, producing the ARF and RMF matrix, and we bin them to have at least 20 counts per bin.
\par In table \ref{tab:extraction} we show the spectral information related to the extracted spectra.
\par Finally, we find no significant evidences of variability between the \textit{Chandra}, \XMM\ and \NuSTAR\ observations.

\begin{table*}[]
\renewcommand{\arraystretch}{1.5}
\center
\caption{Extraction radius in arcsec, spectral counts (source plus background) and signal-to-noise ratio (S/N) for \ngc\ and \eso\ for the three data sets.}
\label{tab:my-table}
\begin{tabular}{cccccccc}
\hline \hline
                & \textbf{} & \multicolumn{3}{c}{\textbf{NGC 3081}}                              & \multicolumn{3}{c}{\textbf{ESO 565-G019}}  \\ \hline
                &           & Extraction radius & Counts      & \multicolumn{1}{c|}{S/N}         & Extraction radius & Counts    & S/N        \\ \hline
\cha\   &           & 2"                & 2623 ± 51   & \multicolumn{1}{c|}{51.0 ± 0.1}  & 2"                & 123 ± 11  & 11.1 ± 0.1 \\
\XMM\     & pn        & 35"               & 3187 ± 56   & \multicolumn{1}{c|}{54.4 ± 0.2}  & 35"               & 742 ± 27  & 24.0 ± 0.5 \\
\textbf{}       & MOS1      & 40"               & 3234 ± 57   & \multicolumn{1}{c|}{54.8 ± 0.2}  & 40"               & 856 ± 29  & 25.6 ± 0.5 \\
\textbf{}       & MOS2      & 45"               & 8788 ± 94   & \multicolumn{1}{c|}{91.1 ± 0.2}  & 45"               & 2147 ± 46 & 42.9 ± 0.3 \\
\NuSTAR\  & FPMA      & 60"               & 16960 ± 130 & \multicolumn{1}{c|}{125.7 ± 0.2} & 40"               & 1067 ± 33 & 26.9 ± 0.6 \\
          & FPMB      & 60"               & 16820 ± 130 & \multicolumn{1}{c|}{125.4 ± 0.2} & 40"               & 1443 ± 38 & 31.4 ± 0.7 \\ \hline
\end{tabular}
\label{tab:extraction}
\end{table*}


%
%
\section{Spectral Models}\label{sec:spectral}
In the following sections we describe the different models used in the spectral analysis. To perform a more physical and detailed analysis of the X-ray spectra, with respect to the classical phenomenological analysis, it is possible to use \textit{physically motivated} models, such as \MYTorus\ and \borus\ (Section \ref{section:MYTorus} and \ref{section:borus}). Both models describe the reprocessing material (i.e., the obscuring torus) in a physical way, using Monte-Carlo simulations. Also, these models allow us to calculate the intrinsic column density of the torus and, in the case of the \texttt{borus02} model, the covering factor, which corresponds to the torus opening angle.


\subsection{\MYTorus}\label{section:MYTorus}
 In this section we will discuss the main properties and the use of the \MYTorus\ model \citep{murphy2009x}. 
The \MYTorus\ model was developed to be used in the \XSPEC\ \citep[][]{arnaud1996xspec} environment as a combination of additive and multiplicative tables, which represent different components of the nuclear emission. These components are:the zeroth-order emission component (MYTZ), the scattered continuum (MYTS) and the iron line emission component (MYTL).
\MYTorus\ models the observed spectrum taking into account the absorbed and the scattered component of the emission, modeling also the presence of fluorescent Fe K$\alpha$ and K$\beta$ emission lines at $6.4$ keV and $7.06$ keV respectively, which are thought to be almost ubiquitous in heavily obscured AGN spectra.


The \MYTorus\ model can be used in two different settings, namely the \textit{coupled} and the \textit{de-coupled} configuration.
In the first mode, the column density and the inclination angle of the three components (MYTZ, MYTS and MYTL) are tied together. Thus, in this configuration, all the components are produced in the same medium.  




The \MYTorus\ model simulates the interaction between input spectrum photons and the obscuring material.  
The circumnuclear environment is simulated as the classical doughnut-like and azimuthally symmetric structure. The distance from the black hole to the center of the torus section is indicated as $c$, and $a$ is the radius of the section.
\par $\theta_{obs}$ is the inclination angle: the angle between the torus symmetry axis and the observer line of sight (l.o.s.). It can vary in the range [0$^{\circ}$ - 90$^{\circ}$], allowing to reproduce both the face-on ($\theta_{obs}$=0$^{\circ}$, i.e., the observer looks directly at the nucleus) and the edge-on ($\theta_{obs}$=90$^{\circ}$,i.e., the observer line of sight intercepts the torus equator).

The torus half-opening angle, which represents the fraction of the sky as seen from the center, is defined as $\alpha = [(\pi-\psi)/2]$=60$^{\circ}$ (with $\psi$ being the angle subtended by the internal surface of the torus) corresponding to a covering factor C$_{TOR}=0.5$. The fixed value for the covering factor is linked to the assumptions made on the fraction of obscured AGN with respect to the unobscured ones. Finally, N$_H$ is the equatorial column density (i.e. the column density through the torus diameter); the line of sight column density can be computed as: $$N_{H, z} = N_H \left[1-\left( \frac{c}{a} \right)^2 \cos^2 \theta_{obs}\right] $$

The first component in the \texttt{MYTorus} model is the so-called zeroth-order continuum or direct component. This component represents the photons escaping the absorbing medium (i.e., the torus) without being absorbed or scattered.

The second component is the scattered or reprocessed continuum, which represents the photons that escape the medium after being scattered one or more times. 
\par 
The interaction is via Compton-scattering, thus the energy of the photon after the scattering will be lower with respect to the input photon's energy. The second component is responsible for the production of the feature observed at $\sim$ 30 keV (i.e. the \textit{Compton hump}). Moreover, in the \MYTorus\ model the termination energy of the scattered component is variable between 160 keV and 500 keV (in our analysis we use a table with intrinsic continuum extending up to 500 keV). The value of the cut-off energy has been chosen to be consistent with previous similar works \citep[see e.g.,][]{Marchesi2017Nustar1, Zhao2019Nustar2, Zhao2019Nustar4}. Moreover, recent works \citep[e.g.,][]{balokovic2020cutoff} show that it is a reasonable value for the extension of the continuum. 

The last component is the fluorescent emission. It takes into account the possibility to have fluorescent emission iron lines, produced in the reprocessing medium.
The emission lines that \MYTorus\ models are the Fe K$\alpha$ and K$\beta$ only.
\par On the one hand, the line photons that escape after being produced by the fluorescence process, constitute the zeroth-order fluorescent emission component. On the other hand, if these photons interact with the reprocessor by scattering processes, they can contribute to form the \textit{Compton shoulder}.

\subsection{\MYTorus\ in “de-coupled" configuration}

The \MYTorus\ model in “coupled" configuration allows the inclination angle to vary, but does not permit a variation in the column density and in the geometrical properties of the different components. In this way it is not possible to properly characterize a clumpy torus structure. To overcome this problem, as described by \cite{yaqoob2012nature}, it is possible to decouple the \MYTorus\ components by fixing the zeroth-order continuum inclination angle to 90$^{\circ}$, generating a pure line of sight component. Then, the column density of the scattered component can be untied with respect to the one of the direct continuum. In this way, the direct continuum column density represents the line of sight column density, whereas the scattered component column density represents the ``global average" column density. Thus, the ratio between the ``global average" column density and the line of sight column density represents a measure of the patchiness (or clumpiness) of the obscuring material: a ratio $\not=1$ will then suggest a scenario in which the column density along the l.o.s. is higher (or lower) than the average column density of the torus, meaning that the structure could likely be clumpy rather than smooth. Following \cite{yaqoob2015compton}, we then fix the inclination angle of the scattered and fluorescent line components to be either $\theta_{S=L}$=90$^{\circ}$ or $\theta_{S=L}$=90$^{\circ}$, reproducing an edge-on and face-on geometry. Using the “de-coupled" mode, it is possible to take into account a scenario in which the different \MYTorus\ components are produced with different interactions of the nuclear emission with the reprocessor.
Using $\theta_{S=L}$=0$^{\circ}$ we model a scenario in which the emission is dominated by reflection in the far-side of the torus; if $\theta_{S=L}$=90$^{\circ}$, the emission is dominated by a near-side Compton-scattering.

\subsection{BORUS02}\label{section:borus}

Finally, we use the \borus\ model (table \texttt{borus02\_v170323a.fits}), developed by \cite{balokovic2018new} as an improvement of the \bntorus\ model \citep{brightman2011xmm}. The \borus\ model is based on grids of spectral templates obtained using Monte-Carlo simulations of radiative transfer through a neutral spherical torus with polar cut-outs. \par The strength of this model lies in the possibility to fit the spectral data having as free parameters the average column density of the torus and its covering factor. The computation of the covering factor it is not possible by using \MYTorus\, even in its “de-coupled" configuration. Due to the longer variability timescales (years), the average column density represents a more reliable parameter to characterize the thickness of an AGN in respect to the N$_{H,los}$, which variability has shown to be of the order of days and weeks, due to the movement of clouds through the line of sight \citep[see e.g.,][]{risaliti2002ubiquitous, ricci2016ic}.
\par Despite this, a proper characterization of the covering factor is not simple. It can be affected by accretion or feedback phenomena taking place nearby the torus \citep[e.g.,][]{heckman2014coevolution, netzer2015revisiting}; it can depend on the luminosity \citep[e.g.,][]{assef2013mid} as well as on the Eddington ratio \citep[e.g.,][]{ricci2017bat}, and these dependencies could vary with redshift \citep[e.g.,][]{buchner2015obscuration}. In this perspective, the \borus\ model represents an updated tool to compute the covering factor. \borus\ model is composed of a single additive table (instead of the three tables of \MYTorus\ ) which takes into account the reprocessed emission component, that is similar to the \MYTorus\ “reprocessed component", and the fluorescent line emission component, including K$\alpha$ and K$\beta$ lines.
\par The main parameters of the \borus\ model have the following possible values: the covering factor ranges from $0.1$ to $1$, corresponding to a torus opening angle between 84$^{\circ}$ and 0$^{\circ}$; the inclination angle is in the range [18$^{\circ}$-87$^{\circ}$]. Also, the cut-off energy is a parameter of the model and we fix it to be 500 keV, for consistency with \texttt{MYTorus}.  Finally, the iron abundance is also a free parameter, but we fix it to 1, for consistency with the \texttt{MYTorus} analysis.


\borus\ does not include the l.o.s. absorption at the redshift of the source, which we model with combination of \XSPEC\ components $zphabs*cabs$ in order to take into account l.o.s. absorption and the losses out of the l.o.s. due to Compton scattering. The primary power law emission is represented by \texttt{cutoffpl1} that is multiplied by the previous expression; it is characterized by a photon index, cut-off energy and normalization, that must be tied to those of \texttt{borus02}. Also, to properly describe the l.o.s. column density, the $nH$ parameter of \texttt{cabs} and \texttt{zphabs} must be tied together. The soft emission component and the emission lines are included, when needed, as described for the \MYTorus\ modelling.
\section{Spectral analysis}\label{analysis}
In this Section we present the spectral analysis of \ngc\ and \eso\ .
Since the background contribution dominates at energies higher than 70 keV and 40 keV, we analyze the spectra up to these energies.
In order to obtain a physically detailed description of the observed emission, we carried out the analysis using the \MYTorus\ and \borus\ physically motivated models. We also add a thermal component (\texttt{mekal}), to reproduce the emission at soft energies, and Gaussian lines at energies $\sim$ 0.92, 1.31 and 1.80 keV, corresponding to Ne IX, Mg XI and Si XIII.

\subsection{NGC 3081}

\subsubsection{\MYTorus\ model}
We use \MYTorus\ in both the “coupled" and “de-coupled" configuration (the last one in the edge-on and face-on mode).
\par
\begin{figure}[]
\centering
{\includegraphics[width=.47\textwidth]{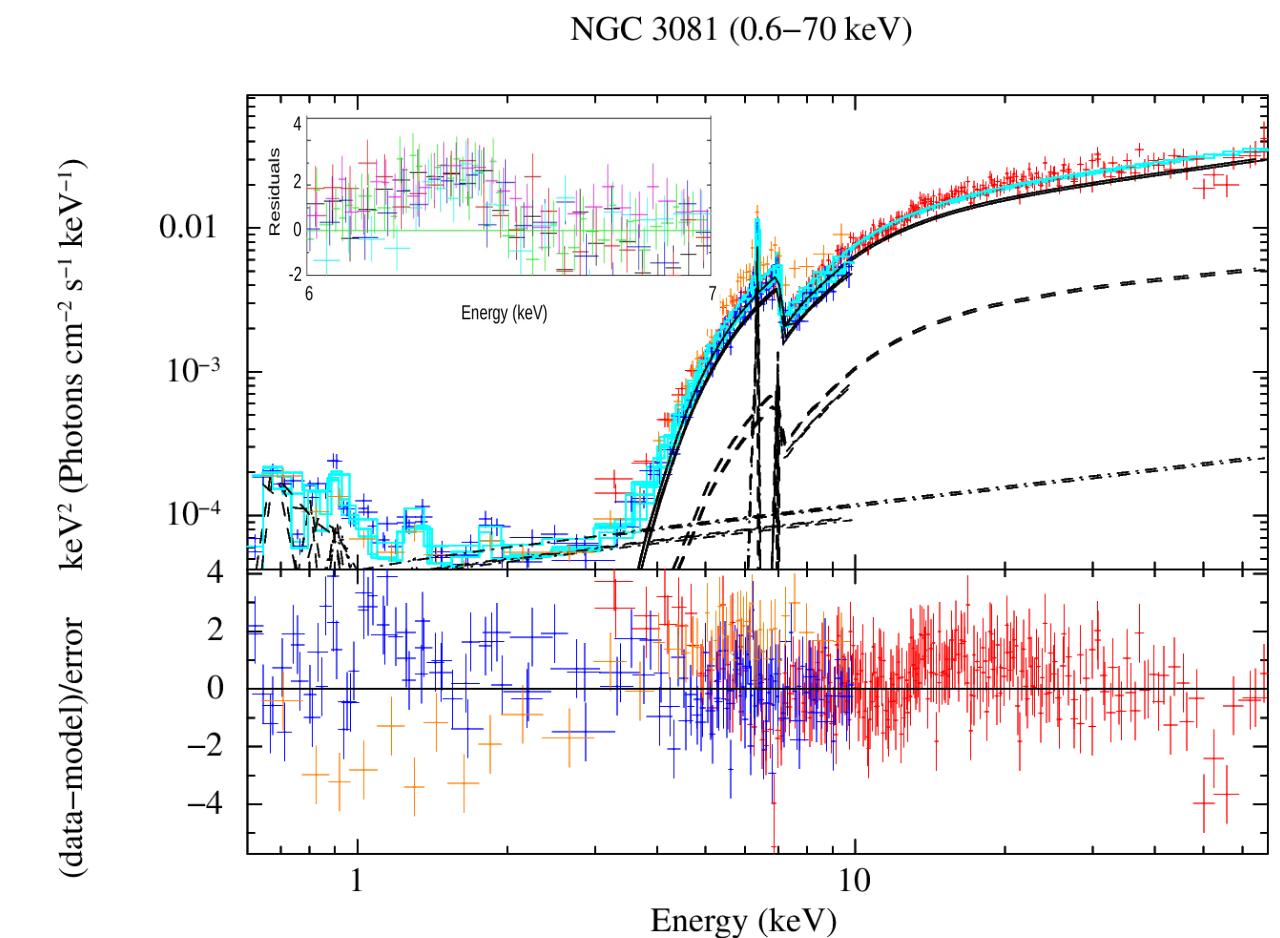}}
\caption{\small{Unfolded \textit{Chandra} (orange), \textit{XMM-Newton} (blue) and \textit{NuSTAR} (red) 0.6--70 keV combined spectrum of NGC 3081 modeled with \texttt{MYTorus} in the “coupled" configuration. The cyan solid line represents the best-fit model, while the individual components, MYTZ, MYTS, MYTL and the second power law, are reported as a black solid line, dashed lines and dash dotted line, respectively. Finally, the mekal component is plotted as a dashed line. In the top left corner the residuals for the iron K$\alpha$ line in the best-fit model without the line component are shown.}}
\label{fig:spec_all_ngc_mytorus_coupled}
\end{figure}

 The best fit model consists of the three \texttt{MYTorus} components, the second power law, the \textit{mekal} component (to model the soft thermal emission) and the emission lines. Moreover, we included other two constants to the models, $A_S$ and $A_L$, to take into account the possible different normalizations of the other two components with respect to the zeroth-order continuum:

\begin{equation}
\label{eq:powerlaw}
\begin{aligned}
Model~NGC\_A =pha*(zpo1*MYTZ + A_S*MYTS +\\
A_L*MYTL + f_s*zpo2+mekal+3*zgauss
\end{aligned}
\end{equation}
\\
The photon index is $\Gamma_{MYT,c,B}=1.59_{-0.03}^{+0.03}$. The column density, which is $N_{H,eq}=0.62_{-0.02}^{+0.02}\times10^{24}$ cm$^{-2}$, is below the Compton-thick threshold. We also fit our data leaving the inclination angle free to vary and we find no significant improvement in the fit statistic ($\chi^2/d.o.f.=1773/1403$). Therefore, we fix the inclination angle to 90$^{\circ}$. The best-fit model and the spectrum are reported in Figure \ref{fig:spec_all_ngc_mytorus_coupled}. Also, we show the residuals for the iron K$\alpha$ line in the best-fit model without the line component. As it can be seen, the line component is required to improve the fit statistic.

\vspace{5mm}
\par In order to reach a more complete knowledge of the geometrical properties of the NGC 3081 torus, we fit the 0.6--70 keV spectrum with \MYTorus\ in the “de-coupled" mode. The photon indices we obtain are $\Gamma_{\theta,S=90}=1.81_{-0.06}^{+0.07}$ and $\Gamma_{\theta,S=0}=1.75_{-0.05}^{+0.03}$, for the edge-on and face-on modes, respectively. Also the l.o.s. column densities are: $N_{H,z}=0.60_{-0.02}^{+0.02}\times 10^{24}$ cm$^{-2}$ and $N_{H,S}=1.59_{-0.17}^{+0.19}\times 10^{24}$ cm$^{-2}$ for the edge-on mode and $N_{H,z}=0.66_{-0.02}^{+0.02}\times 10^{24}$ cm$^{-2}$ and $N_{H,S}=3.00_{-0.68}^{+0.69}\times 10^{24}$ cm$^{-2}$ for the face-on mode.
In both modes the photon index results to be steeper than the one found in the analysis with the “coupled" configuration (see Table \ref{tab:ngc}); also the l.o.s. column densities are lower than the ``global average" ones.
The $\chi^2$ statistics favors the edge-on configuration, with a $\chi^2/d.o.f. = 1723/1403$.  We show in Figure \ref{fig:ngc_final_spectra} (left panel) the unfolded spectrum and the \MYTorus\ model in the edge-on mode.

\subsubsection{\borus\ model}
\begin{figure*} 
\begin{minipage}[b]{.5\textwidth}
\centering
\includegraphics[width=\textwidth]{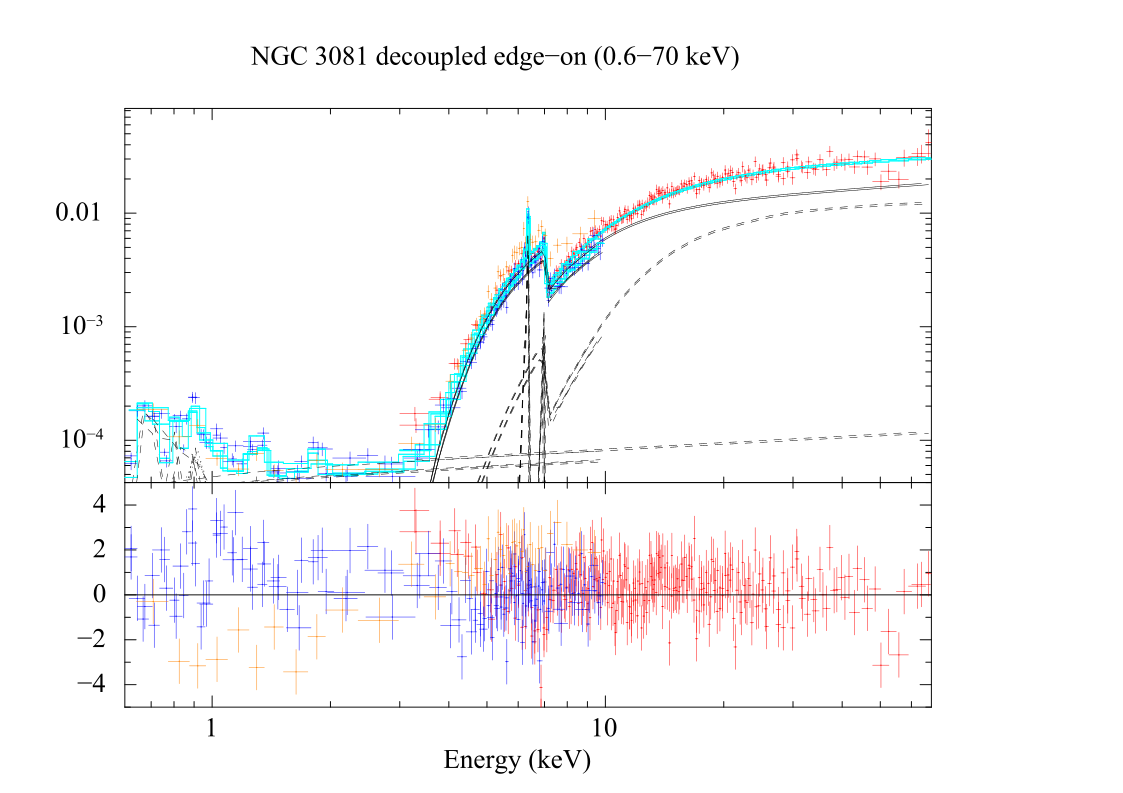}
\end{minipage}
\begin{minipage}[b]{.5\textwidth}
\includegraphics[width=\textwidth]{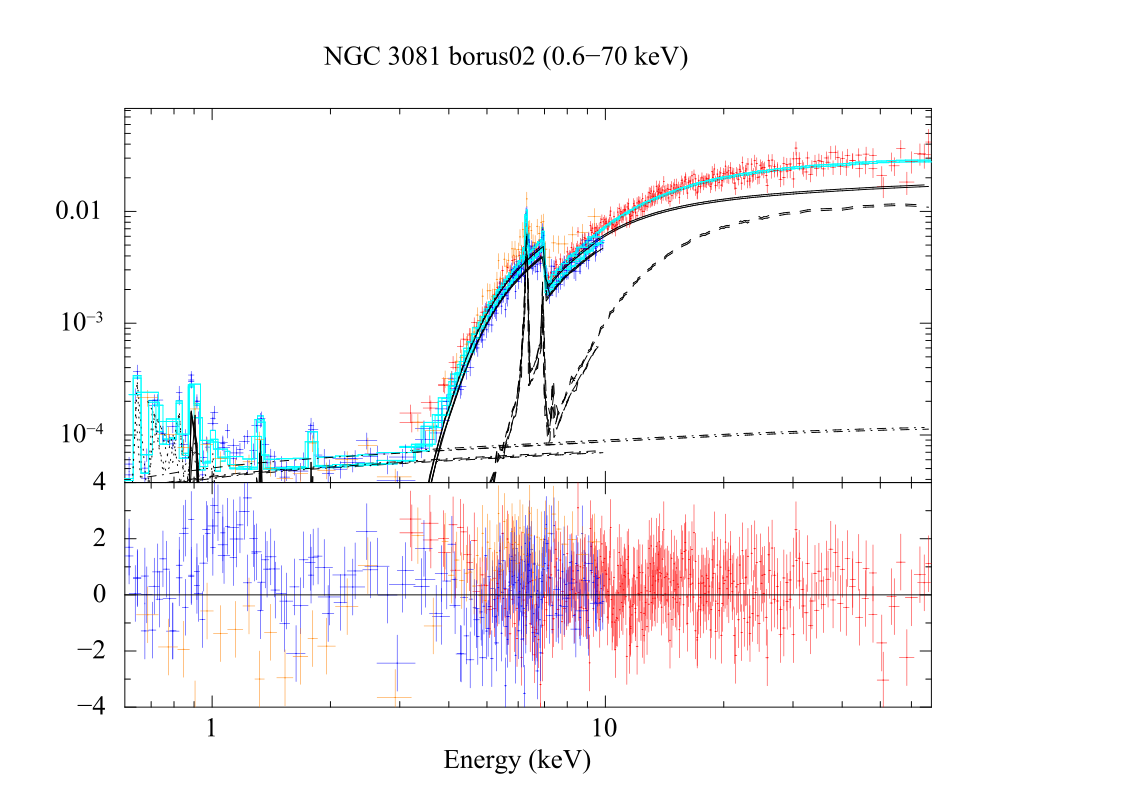}
\end{minipage}
\caption{Unfolded \textit{Chandra} (orange), \textit{XMM-Newton} (blue) and \textit{NuSTAR} (red) $0.6-70$ keV combined spectrum of NGC 3081 modeled with \texttt{MYTorus} in the “de-coupled" configuration in the edge-on mode (left) and \texttt{borus02} (right). In the case of \MYTorus\ the best fit model (cyan solid line) and the individual components are plotted as in Figure \ref{fig:spec_all_ngc_mytorus_coupled}. The \borus\ and the emission line components are plotted as a dashed line, the first power law is plotted as a solid line and the second power law as a dot-dashed line.}
\label{fig:ngc_final_spectra}
\end{figure*}

Finally, we model the $0.6-70$ keV combined spectrum with the \borus\ model. In addition to the main emission component and the reprocessed component, we also included the second power law and the thermal component to model the contribution in the soft part of the spectrum. Also, we add the three emission lines previously described.

\begin{equation}
\label{eq:modelngcB}
\begin{aligned}
Model~NGC\_B =pha*(borus02 + zpha*cabs*cutoffpl1\\
+ f_s*cutoffpl2+mekal+3*zgauss)
\end{aligned}
\end{equation}
\\
The best-fit model has a $\chi^2/d.o.f.=1753/1403$; the photon index is $\Gamma=1.80_{-0.04}^{+0.06}$; the l.o.s. column density is $N_{H,z}=0.61_{-0.02}^{+0.02}\times10^{24}$ cm$^{-2}$ and the average column density $N_{H,S}=1.51_{-0.10}^{+0.11}\times10^{24}$ cm$^{-2}$ is consistent with the torus being Compton-thick, as we also found using \MYTorus\ in the edge-on configuration. In Figure \ref{fig:ngc_final_spectra} (right panel) we show the 0.6--70 keV spectrum. The \borus\ model allows us to leave the covering factor as a free parameter; in the case of \ngc\ we obtain C$_{TOR}=0.73_{-0.10}^{+0.09}$.

\begingroup
\renewcommand{\arraystretch}{1.5}
\begin{table*}[]
\centering
\caption{Summary table of the spectral results obtained with \MYTorus\ (coupled and decoupled) and \borus\ applied to NGC 3081 data.}
\label{tab:ngc}
\begin{tabular}{ccccc}
\hline \hline
                & \multicolumn{4}{c}{\textbf{NGC 3081}}                                                                                                                                                                                   \\ \hline
                & \textbf{\begin{tabular}[c]{@{}c@{}}MYTorus\\ “coupled"\end{tabular}} & \multicolumn{2}{c}{\textbf{\begin{tabular}[c]{@{}c@{}}MYTorus\\ “de-coupled"\end{tabular}}} & \textbf{borus02}        \\
                & \textbf{}                                                            & \textbf{edge-on}                             & \textbf{face-on}                             & \textbf{}               \\ \hline
$\chi^2/d.o.f.$ & $1774/1404$                                                          & $1723/1403$                                  & $1747/1403$                                  & $1753/1403$             \\
$\Gamma$       & $1.59_{-0.03}^{+0.03}$                                               & $1.81_{-0.06}^{+0.07}$                       & $1.75_{-0.05}^{+0.03}$                       & $1.80_{-0.04}^{+0.06}$  \\
$N_{H,eq}\footnote{Equatorial column density in the \MYTorus\ model, in the “coupled" configuration, in unit of 10$^{24}$ cm$^{-2}$.}$      & $0.62_{-0.02}^{+0.02}$                                               & $\cdot~\cdot~\cdot$                          & $\cdot~\cdot~\cdot$                          & $\cdot~\cdot~\cdot$     \\
$N_{H,z}\footnote{Column density along the l.o.s. in unit of 10$^{24}$ cm$^{-2}$.}$       & $\cdot~\cdot~\cdot$                                                  & $0.60_{-0.02}^{+0.02}$                       & $0.66_{-0.02}^{+0.02}$                       & $0.61_{-0.02}^{+0.02}$  \\
$N_{H,S}\footnote{“Global average" column density in unit of 10$^{24}$ cm$^{-2}$.}$       & $\cdot~\cdot~\cdot$                                                  & $1.59_{-0.17}^{+0.19}$                       & $3.00_{-0.68}^{+0.69}$                       & $1.51_{-0.10}^{+0.11}$  \\
$A_{S}=A_{L}\footnote{Normalization between the reprocessed \MYTorus\ component and the zeroth-order continuum.}$   & $0.86_{-0.12}^{+0.13}$                                               & $2.23_{-0.37}^{+0.46}$                       & $0.42_{-0.05}^{+0.05}$                       & $\cdot~\cdot~\cdot$     \\
$\theta_{obs}\footnote{Torus inclination angle in degrees.}$  & $90^f\footnote{The f indicates that a parameter is fixed.}$                                                               & $90^f$                                       & $0^f$                                       & $\cdot~\cdot~\cdot$     \\
$f_{s}\footnote{Fraction of the scattered component.}$         & $5.63_{-0.56}^{+0.62}$                                               & $4.36_{-0.62}^{+0.57}$                       & $3.17_{-0.34}^{+0.53}$                       & $4.51_{-0.55}^{+0.44}$  \\
$EW\footnote{Equivalent width of the K$\alpha$ Iron line in unit of keV.}$            & $0.181_{-0.003}^{+0.010}$                                            & $0.185_{-0.007}^{+0.007}$                    & $0.185_{-0.007}^{+0.007}$                    & $\cdot~\cdot~\cdot$     \\
$C_{TOR}\footnote{Covering factor of the torus, in the \borus\ model.}$       & $\cdot~\cdot~\cdot$                                                  & $\cdot~\cdot~\cdot$                          & $\cdot~\cdot~\cdot$                          & $0.73_{-0.10}^{+0.09}$  \\
$kT\footnote{Temperature of the thermal component in keV.}$            & $0.26_{-0.01}^{+0.02}$                                               & $0.25_{-0.02}^{+0.02}$                       & $0.26_{-0.02}^{+0.02}$                       & $0.26_{-0.02}^{+0.02}$  \\
$F_{2-10~keV}\footnote{2--10 keV flux in unit of $10^{-12}$ erg s$^{-1}$ cm$^{-2}$.}$  & $4.50_{-0.13}^{+0.11}$                                               & $4.40_{-0.19}^{+0.10}$                       & $4.45_{-0.17}^{+0.11}$                       & $4.38_{-0.14}^{+0.10}$  \\
$F_{10-40~keV}\footnote{10--40 keV flux in unit of $10^{-12}$ erg s$^{-1}$ cm$^{-2}$.}$ & $4.12_{-0.08}^{+0.07}$                                               & $4.19_{-0.13}^{+0.05}$                       & $4.16_{-0.28}^{+0.03}$                       & $4.18_{-0.08}^{+0.10}$  \\
$log(L_{2-10~keV})\footnote{2--10 keV intrinsic luminosity in unit of erg s$^{-1}$.}$  & $41.73_{-0.01}^{+0.01}$                                              & $41.74_{-0.05}^{+0.05}$                      & $41.75_{-0.06}^{+0.06}$                      & $42.78_{-0.01}^{+0.01}$ \\
$log(L_{10-40~keV})\footnote{10--40 keV intrinsic luminosity in unit of erg s$^{-1}$.}$  & $42.56_{-0.02}^{+0.02}$                                              & $42.57_{-0.01}^{+0.01}$                      & $42.66_{-0.07}^{+0.07}$                      & $42.83_{-0.01}^{+0.01}$            \\ \hline
\end{tabular}
\end{table*}
\endgroup

\subsubsection{Summary of the NGC 3081 spectral analysis results}
We have analyzed the NGC 3081 0.6--70 keV \textit{Chandra}, \XMM\ and \NuSTAR\ combined spectra, which have a high count statistics (N$_C=53.6$ k-counts, considering all instruments). The spectral analysis has been done with both \MYTorus\ and \borus\ physically motivated models. The best fit model, in terms of lowest $\chi_{\nu}^2$, is the \MYTorus\ one, used in the “de-coupled" mode in the edge-on configuration, with a reduced statistics $\chi_{\nu}^2=\chi^2/d.o.f. = 1723/1403$ (the \borus\ best-fit model has $\chi_{\nu}^2=\chi^2/d.o.f. = 1753/1403$).

The main goal of the present work is the classification of the sources under investigation through the determination of the column density of the obscuring material (i.e., the torus). The \MYTorus\ model (as well as the \borus\ model), in its “de-coupled" mode, suits very well for this purpose, because it allows to distinguish between the l.o.s. column density (N$_{H,z}$) and the “global average" column density of the torus (N$_{H,S}$). From the best-fit model we find that along the l.o.s. NGC 3081 results to be Compton-thin (log(N$_{H,z}$/[cm$^{-2}$])$= 23.78_{-0.02}^{+0.01}$). However, the average column density of its torus (log(N$_{H,S}$/[cm$^{-2}$])$= 24.20_{-0.05}^{+0.05}$) is above the Compton-thick threshold. This scenario is typical of sources that are being observed through a lower density portion of the torus with respect to its average density, suggesting that a patchy or clumpy structure is preferred to the classical smooth, doughnut-like, geometry. Thus, we can affirm that NGC 3081 has a CT torus (at the 90\% confidence level) observed through a Compton-thin portion of the obscuring material. In addition, from \borus\ spectral analysis, which results on the main spectral parameters are consistent with the \MYTorus\ ones, we can obtain the torus covering factor, that is found to be $C_{TOR}=0.73_{-0.10}^{+0.09}$, corresponding to a torus half opening angle of $\sim$ 43$^{\circ}$.
\par Although this analysis was focused on the investigation of the thickness and geometry of the obscuring material, we also study the properties of the soft X-ray emission. We find that contribution to the soft emission comes from several components: the fraction of photon that are scattered, rather than being absorbed, by Compton-thin material is lower than 1\% of the main emission component, consistent with the average value obtained for obscured AGN \citep{marchesi2018compton}; the 0.6--2 keV emission is well fitted by adding a thermal component, originated by presence of diffuse gas in the nuclear region, with a temperature of kT$\sim 0.3$ keV; we also detect several emission lines at energies $\sim$0.92, 1.31 and 1.80 keV, which are expected to arise from the continuum in the case of obscured sources (if the statistics is sufficiently high to detect them), associated to Ne IX, Mg XI and Si XIII \citep[also found in other obscured AGN, related to the ionizing AGN flux, e.g.,][]{brinkman2002soft, piconcelli2011x}.

\par We also compute the mid-IR luminosity following the relation presented by \cite{Asmus2015midir}. The mid-IR luminosity at 12$\mu$m is log(L$_{12\mu m}$)= $43.10_{-0.04}^{+0.04}$ erg s$^{-1}$ using the \texttt{borus02} 2--10 keV luminosity and log(L$_{12\mu m}$)= $42.02_{-0.13}^{+0.12}$ erg s$^{-1}$ with the \MYTorus\ 2--10 keV luminosity. Since the value obtained by \cite{Asmus2015midir} is log(L$_{12\mu m}$)= $42.87_{-0.07}^{+0.07}$ erg s$^{-1}$ it can suggest that the \borus\ model allows us to reach a representation of the intrinsic emission. Finally, we computed the iron K$\alpha$ emission line equivalent width, which is $\sim 0.180$ keV (see Table \ref{tab:ngc}). Although this value is lower than the typical threshold for CT-AGN \citep[$\sim$1 keV, see, e.g.,][]{koss2016new}, there is evidence of similar sources in previous literature works \citep[e.g.,][]{Marchesi2017Nustar1}.
\subsection{ESO 565-G019}

\subsubsection{\MYTorus\ model}
\begin{figure}[]
\centering
{\includegraphics[width=.47\textwidth]{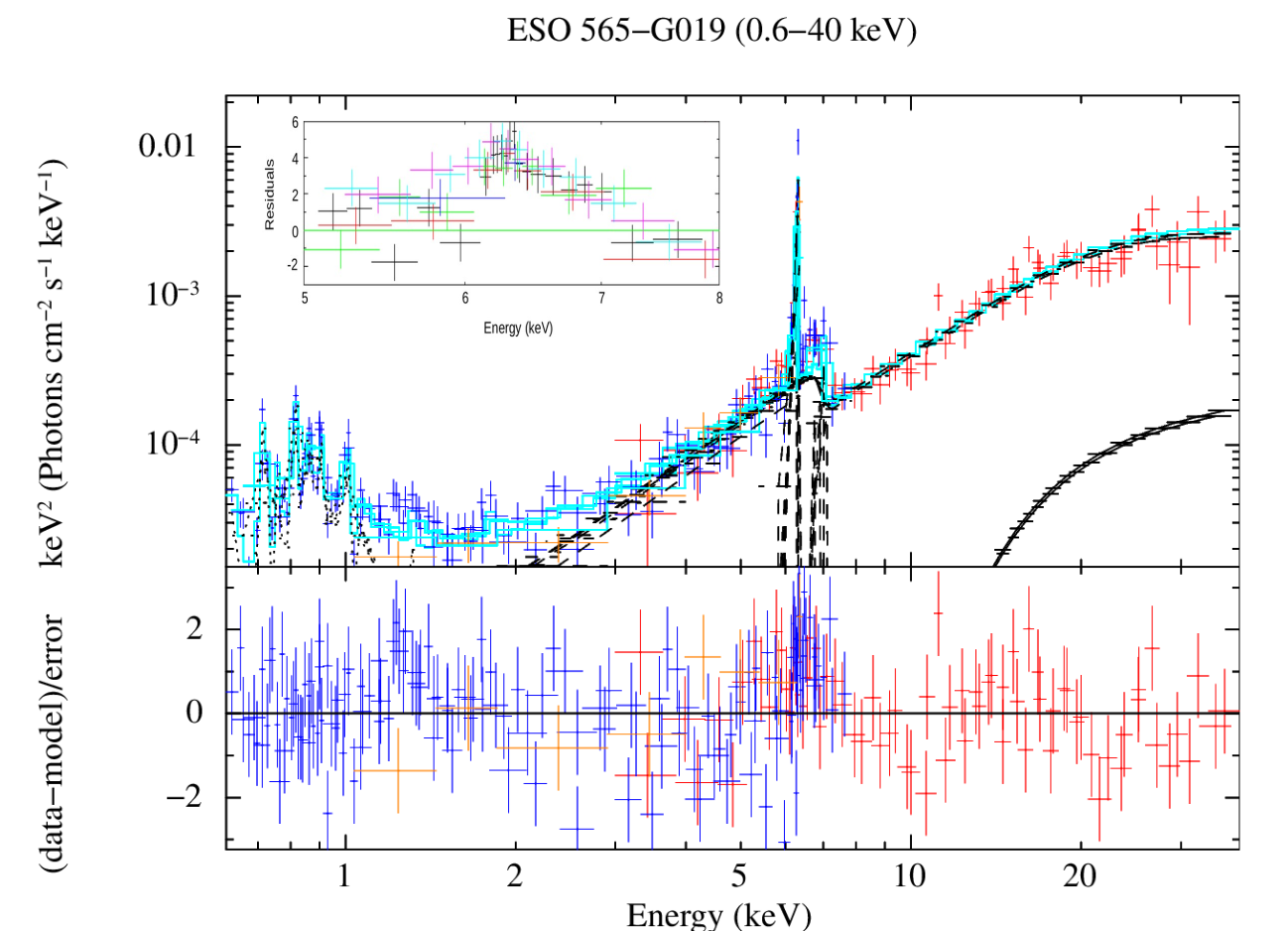}}
\caption{\small{Unfolded \textit{Chandra} (orange), \textit{XMM-Newton} (blue) and \textit{NuSTAR} (red) 0.6--40 keV combined spectrum of \eso\ modeled with \texttt{MYTorus} in the “coupled" configuration. The cyan solid line represents the best-fit model, while the individual components, MYTZ, MYTS, MYTL and the second power law, are reported as a black solid line, dashed lines and dash dotted line, respectively. Finally, the mekal component is plotted as dashed line. In the top left corner the residuals for the iron K$\alpha$ line in the best-fit model without the line component are shown.}}
\label{fig:spec_all_eso_mytorus_coupled}
\end{figure}

We firstly analyze the \eso\ 0.6--40 keV spectrum with \MYTorus\ in its “coupled" configuration, using the three \MYTorus\ components plus the second power law and the thermal component to describe the soft part of the spectrum:

\begin{equation}
\label{eq:powerlaw}
\begin{aligned}
model~ESO\_A=pha(zpo1*MYTZ + A_S*MYTS+\\
A_L*MYTL+mekal+f_{s}*zpo2)
\end{aligned}
\end{equation}
\\
In this case, leaving the inclination angle free to vary, it is not possible to obtain a statistically acceptable solution for the fit ($\chi^2 > 2$). To this purpose, we tried two different configurations: one with the inclination angle fixed to 90$^{\circ}$, and the other with $\theta_{obs}$=65$^{\circ}$ (i.e., we are observing through the brink of the torus). We report the results of the spectral fitting in Table \ref{tab:eso}: both the photon index and the column density are very different in the two models, in particular we obtain $N_{H,MYT,c,65}=10_{-1.29}^{+0.00*}\times 10^{24}$ cm$^{-2}$ that is the \texttt{MYTorus} upper limit. We also try to fit the data fixing the angle to an intermediate value ($\theta_{obs}$=77$^{\circ}$), but the statistics does not show significant improvement ($\chi_{\nu}^2=297/212$). We report in Figure \ref{fig:spec_all_eso_mytorus_coupled} the 0.6--40\,keV spectrum of ESO 565-G019 fitted with MyTorus coupled with $\theta$=65$^{\circ}$.
\par These fitting issues may be due to the limitations of the “coupled" mode, thus, using \texttt{MYTorus} in its “de-coupled" configuration, we expect to achieve a more physical description of the circumnuclear region for ESO 565-G019.

We then model the spectrum using the \texttt{MYTorus} model in the edge-on and face-on modes of the “de-coupled" configuration and adding both the second power law and the thermal component to reproduce the soft emission. In the edge-on configuration the photon index is $\Gamma_{\theta,S=90}=1.56_{-0.16*}^{+0.16}$; in the face-on mode is instead steeper $\Gamma_{\theta,S=90}=2.22_{-0.08}^{+0.12}$. The l.o.s. column densities are $N_{H,z}=2.96_{-0.38}^{+0.53}\times10^{24}$ cm$^{-2}$ for the edge-on mode and $N_{H,z}=5.80_{-2.53}^{+4.20*}\times10^{24}$ cm$^{-2}$ in the face-on configuration. The ``global average" column density is $N_{H,S}=0.35_{-0.05}^{+0.07}\times10^{24}$ cm$^{-2}$ in the edge-on configuration and $N_{H,S}=3.30_{-1.80}^{+1.19}\times10^{24}$ cm$^{-2}$ in the other one. For both modes the preferred scenario is the one in which we are observing through a particularly dense region of the torus, which has a lower ``global average" column density. In Table \ref{tab:eso} we report the spectral parameter for the 0.6--40 keV spectra: as it can be seen, the statistically favored scenario is the face-on ones, whose best-fit model with the spectrum is reported in Figure \ref{fig:eso_final_spectra}.

\subsubsection{\borus\ model}
\begin{figure*} 
\begin{minipage}[b]{.5\textwidth}
\centering
\includegraphics[width=\textwidth]{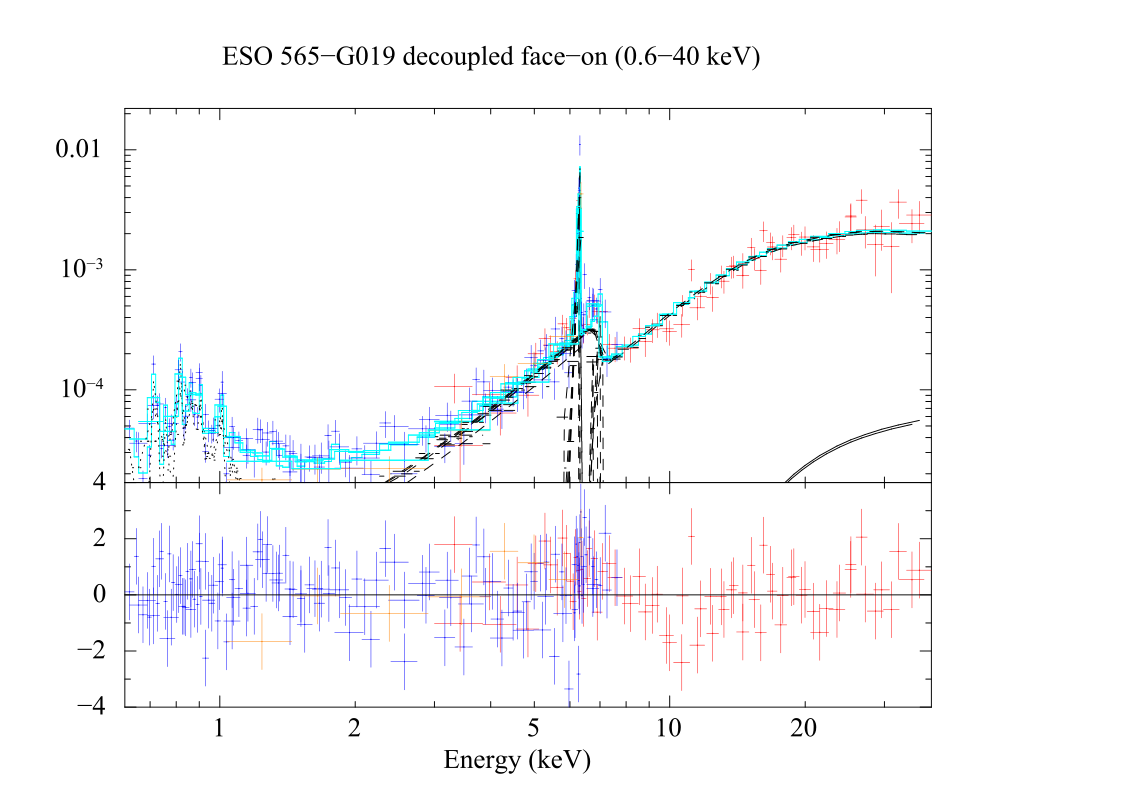}
\end{minipage}
\begin{minipage}[b]{.5\textwidth}
\includegraphics[width=\textwidth]{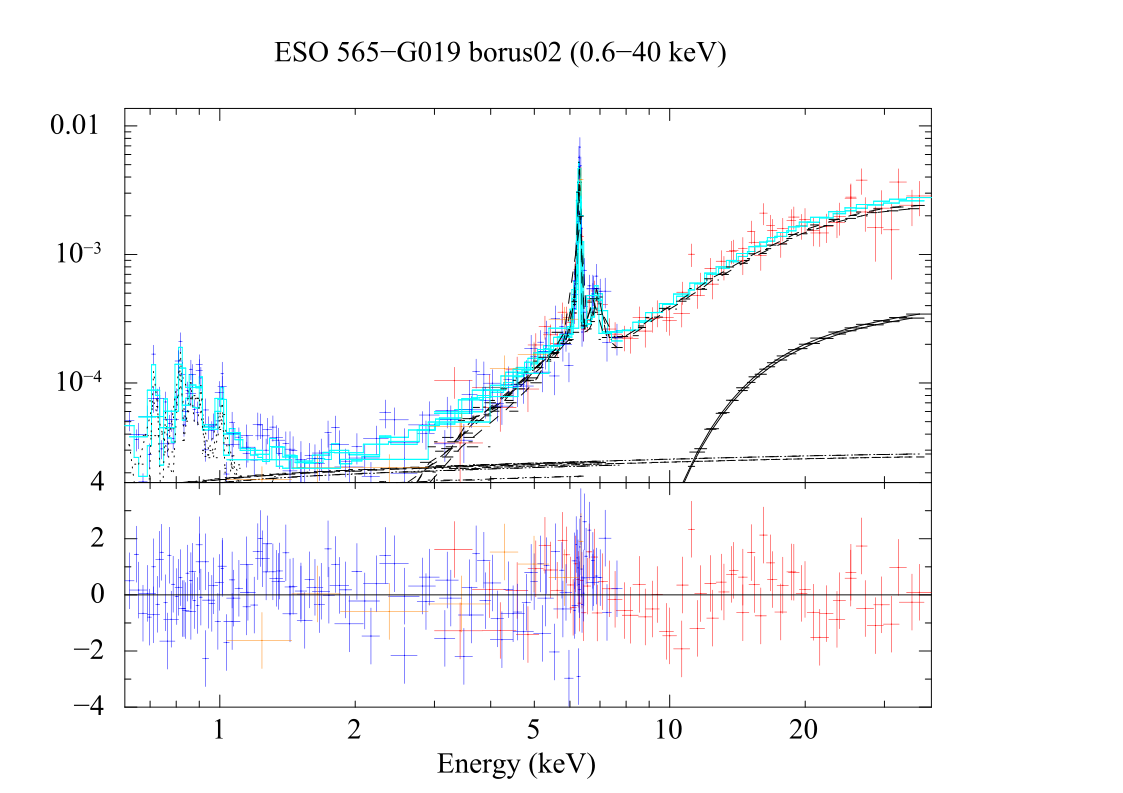}
\end{minipage}
\caption{Unfolded \textit{Chandra} (orange), \textit{XMM-Newton} (blue) and \textit{NuSTAR} 0.6--40 keV combined spectrum of \eso\ modeled with \texttt{MYTorus} in the “de-coupled" configuration in the face-on mode (left) and \borus\ (right). The best fit model and the individual components are plotted as in Figure \ref{fig:spec_all_ngc_mytorus_coupled}, for the \MYTorus\ model. The \borus\ and the emission line components are plotted as a dashed line, the first power law is plotted as a solid line and the second power law as a dot-dashed line.}
\label{fig:eso_final_spectra}
\end{figure*}

We finally analyze the \eso\ spectrum using the \borus\ spectral model. The model consists of the \texttt{borus02} table, the two powerlaws with a cutoff and the \texttt{mekal} component to take into account the soft emission: 

\begin{equation}
\label{eq:powerlaw}
\begin{aligned}
model~ESO\_B=pha*(borus02+zpha*cabs\\
*cutoffpl1+C2*cutoffpl2+mekal)
\end{aligned}
\end{equation}
\\
The best fit model ($\chi^2/d.o.f. = 248/209$) is characterized by a photon index $\Gamma=1.75_{-0.22}^{+0.04}$ and a l.o.s. absorption that is consistent with a Compton-thick scenario $N_{H,l.o.s.}=3.72_{-1.15}^{+*} \times 10^{24}$ cm$^{-2}$. The average column density is also Compton-thick, but unconstrained in its upper bound and the covering factor is $C_{TOR}=0.47_{-0.08}^{+0.18}$, We show the results in Table \ref{tab:eso} and the combined 0.6--40 keV spectrum in Figure \ref{fig:eso_final_spectra}.

\begin{table*}[]
\centering
\renewcommand{\arraystretch}{1.5}
\caption{Summary table of the spectral results obtained with \MYTorus\ (coupled and decoupled) and \borus\ applied to ESO 565-G019 data. The parameters are reported as in Table \ref{tab:ngc}.}
\label{tab:eso}
\begin{tabular}{ccccc}
\hline \hline
                & \multicolumn{4}{c}{\textbf{ESO 565-G019}}                                                                                                                                                                                \\ \hline
                & \textbf{\begin{tabular}[c]{@{}c@{}}MYTorus\\ “coupled"\end{tabular}} & \multicolumn{2}{c}{\textbf{\begin{tabular}[c]{@{}c@{}}MYTorus\\ “de-coupled"\end{tabular}}} & \textbf{borus02}        \\
                & \textbf{}                                                            & \textbf{edge-on}                             & \textbf{face-on}                             & \textbf{}               \\ \hline
$\chi^2/d.o.f.$ & $286/212$                                                            & $292/211$                                    & $256/211$                                    & $248/209$               \\
$\Gamma$        & $2.08_{-0.08}^{+0.07}$                                               & $1.56_{-*}^{+0.16}$                          & $2.22_{-0.08}^{+0.12}$                       & $1.75_{-0.22}^{+0.04}$  \\
$N_{H,eq}$      & $5.30_{-0.68}^{+^*}$                                                 & $\cdot~\cdot~\cdot$                          & $\cdot~\cdot~\cdot$                          & $\cdot~\cdot~\cdot$     \\
$N_{H,z}$       & $\cdot~\cdot~\cdot$                                                  & $2.96_{-0.38}^{+0.53}$                       & $5.80_{-2.53}^{+4.20}$                       & $3.00_{-0.69}^{+*}$     \\
$N_{H,S}$       & $\cdot~\cdot~\cdot$                                                  & $0.35_{-0.05}^{+0.07}$                       & $3.30_{-1.80}^{+1.19}$                       & $3.72_{-1.15}^{+*}$     \\
$A_{S}=A_{L}$   & $1^f$                                                                & $1^f$                                        & $1^f$                                        & $\cdot~\cdot~\cdot$     \\
$\theta_{obs}$  & $65^f$                                                               & $90^f$                                       & $0^f$                                       & $\cdot~\cdot~\cdot$     \\
$f_{s}$         & $1.24_{-0.42}^{+0.35}$                                               & $6.61_{-1.53}^{+2.03}$                       & $1.73_{-0.60}^{+1.66}$                       & $9.14_{-4.58}^{+4.09}$  \\
$EW$            & $1.90_{-0.27}^{+*}$                                               & $1.42_{-0.77}^{+0.18}$                    & $1.60_{-0.0.42}^{+*}$                    & $\cdot~\cdot~\cdot$     \\
$C_{TOR}$       & $\cdot~\cdot~\cdot$                                                  & $\cdot~\cdot~\cdot$                          & $\cdot~\cdot~\cdot$                          & $0.47_{-0.08}^{+0.18}$  \\
$kT$            & $0.59_{-0.03}^{+0.03}$                                               & $0.57_{-0.05}^{+0.03}$                       & $0.59_{-0.04}^{+0.03}$                       & $0.59_{-0.03}^{+0.03}$  \\
$F_{2-10~keV}$  & $0.48_{-0.10}^{+24.25}$                                              & $0.52_{-0.10}^{+0.03}$                       & $0.48_{-0.09}^{+16.42}$                      & $0.50_{-0.39}^{+5.07}$  \\
$F_{10-40~keV}$ & $3.89_{-0.07}^{+13.42}$                                              & $3.61_{-0.86}^{+0.30}$                       & $3.44_{-0.18}^{+9.51}$                       & $3.68_{-2.79}^{+2.53}$  \\
$log(L_{2-10~keV})$  & $43.17_{-0.06}^{+0.05}$                                             & $42.87_{-0.03}^{+0.03}$                      & $43.03_{-0.05}^{+0.04}$                                 & $42.48_{-0.63}^{+0.25}$ \\
$log(L_{10-40~keV})$ & $42.22_{-0.07}^{+0.06}$                                             & $42.12_{-0.04}^{+0.03}$                      & $42.04_{-0.06}^{+0.05}$                                 & $42.56_{-0.56}^{+0.24}$ \\ \hline
\end{tabular}
\end{table*}
\vspace{10mm}

\subsubsection{Summary of the \eso\ spectral analysis results}
ESO 565-G019 has lower spectral counts with respect to NGC 3081, being $N_C=5.8$ k-counts. For ESO 565-G019 the best-fit model is the one given by the \texttt{borus02} analysis, with $\chi_{\nu}^2=248/209$. \texttt{borus02} provides an estimate of the average column density of the obscuring torus. It results to be slightly larger than the l.o.s. column density, but still consistent with it (i.e., both have upper values consistent with the upper boundary allowed by the model). Within the uncertainties, the source is Compton-thick at the $>3 \sigma$ confidence level in both the l.o.s. and average column densities. Moreover, we computed the covering factor, which is $C_{TOR}=0.47_{-0.08}^{+0.18}$, that corresponds to an half opening angle of the torus $\sim$ 62$^{\circ}$. The 10-40 keV flux is $4.18_{-0.08}^{+0.10} \times 10^{-12}$ erg s$^{-1}$ cm$^{-2}$, consistent with the {\it Swift}/BAT, but lower than the {\it Suzaku}/HXD \citep[][]{gandhi2013reflection}, suggesting possible long-term flux variability. 
\par The soft emission in ESO 565-G019 has been modeled combining a contribution of the scattered, unabsorbed fraction of the main emission and the thermal emission component with kT$\sim 0.6$ keV. For this AGN, we do not find any statistically significant emission line at soft energies ($E<2$ keV). Finally, we computed the equivalent width of the iron K$\alpha$ emission line. Its value ($\sim 1.60$ keV, see Table \ref{tab:eso}) is beyond the threshold usually adopted to select candidate CT-AGN \citep[EW$>1$ keV, see, e.g.,][]{koss2016new}.


%
%
\section{Discussion and Conclusions}\label{discussion}
\subsection{The advantages of the \NuSTAR\ approach}
We have analyzed the “soft" \cha\ and \XMM\ spectra alone in order to quantify the effect of adding the information from the \NuSTAR\ data. From the 0.6--10 keV analysis of NGC 3081, we obtain a best fit model with $\Gamma = 1.91_{-0.16}^{+0.15}$ and $N_{H,z}=0.74_{-0.07}^{+0.06} \times 10^{24}$ cm$^{-2}$. The spectral slope is in agreement with typical values observed in AGN. However, in order to increase the statistics and properly constrain the spectral parameters (in particular, the photon index and the column density), we have combined the \textit{NuSTAR} data with the “soft" spectrum, obtaining the 0.6--70 keV NGC 3081 spectrum. As expected, the uncertainties on the main spectral parameters significantly decrease: the errors associated to the photon index decrease from $\sim 8\%$ to $\sim 4\%$ and those on the l.o.s. column density decrease from $\sim 8\%$ to $\sim 3\%$. Also, in accordance with \cite{marchesi2018compton}, we find a shift in both the photon index and the l.o.s. column density values. The first is reduced by $\sim 5\%$ \citep[the average decrease in $\Gamma$ value found by][]{marchesi2018compton} is $\sim 13\%$ and the $N_{H,z}$ one decreases by $\sim 19\%$ \citep[the average value in ][]{marchesi2018compton} is $\sim 32\%$; however, several sources in that sample only had low-count statistic \textit{Swift}-XRT coverage in the 0.5--10\,keV band). The smaller errors allow us to break the degeneracy between $\Gamma$ and N$_{H,z}$. In Figure \ref{fig:ctplot} we show the comparison between the 0.6--10 keV and 0.6--70 keV $\Gamma$-N$_{H,z}$ confidence regions. It is clear that, when adding \textit{NuSTAR} data to the 0.6--10 keV spectrum, there is a shift in the spectral parameters to lower values and, also, a significant decrease of their uncertainties; this result highlights the strength of the X-ray broad-band approach to characterize candidate CT-AGN, and the key role played by \NuSTAR\ to achieve this goal.

\subsection{Variability}
\begin{figure*}[]
\begin{minipage}[b]{.63\textwidth}
\centering
{\includegraphics[width=1.\textwidth]{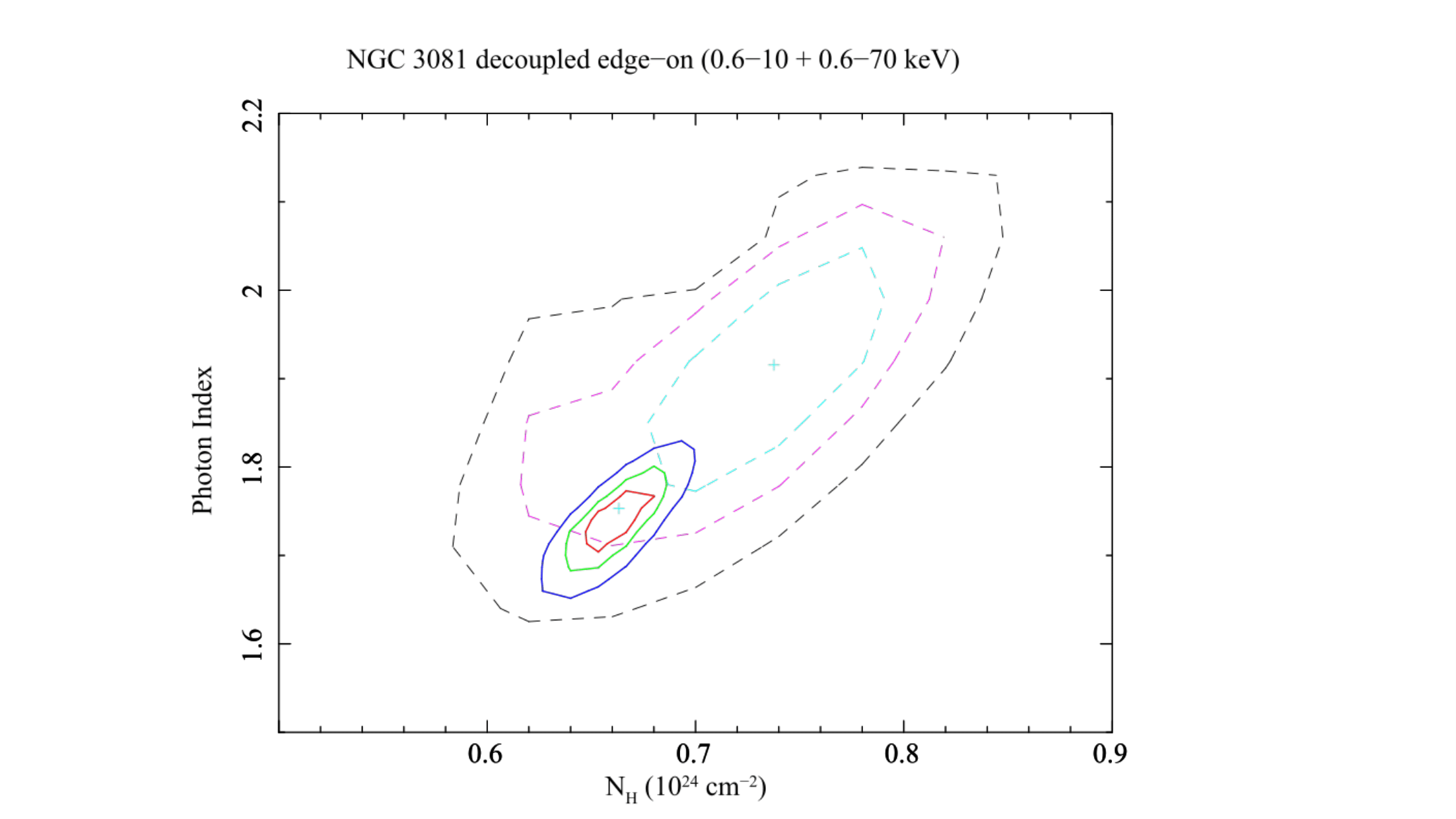}}
\end{minipage}
\hspace{-20mm}
\begin{minipage}[b]{.5\textwidth}
\centering
{\includegraphics[width=1.\textwidth]{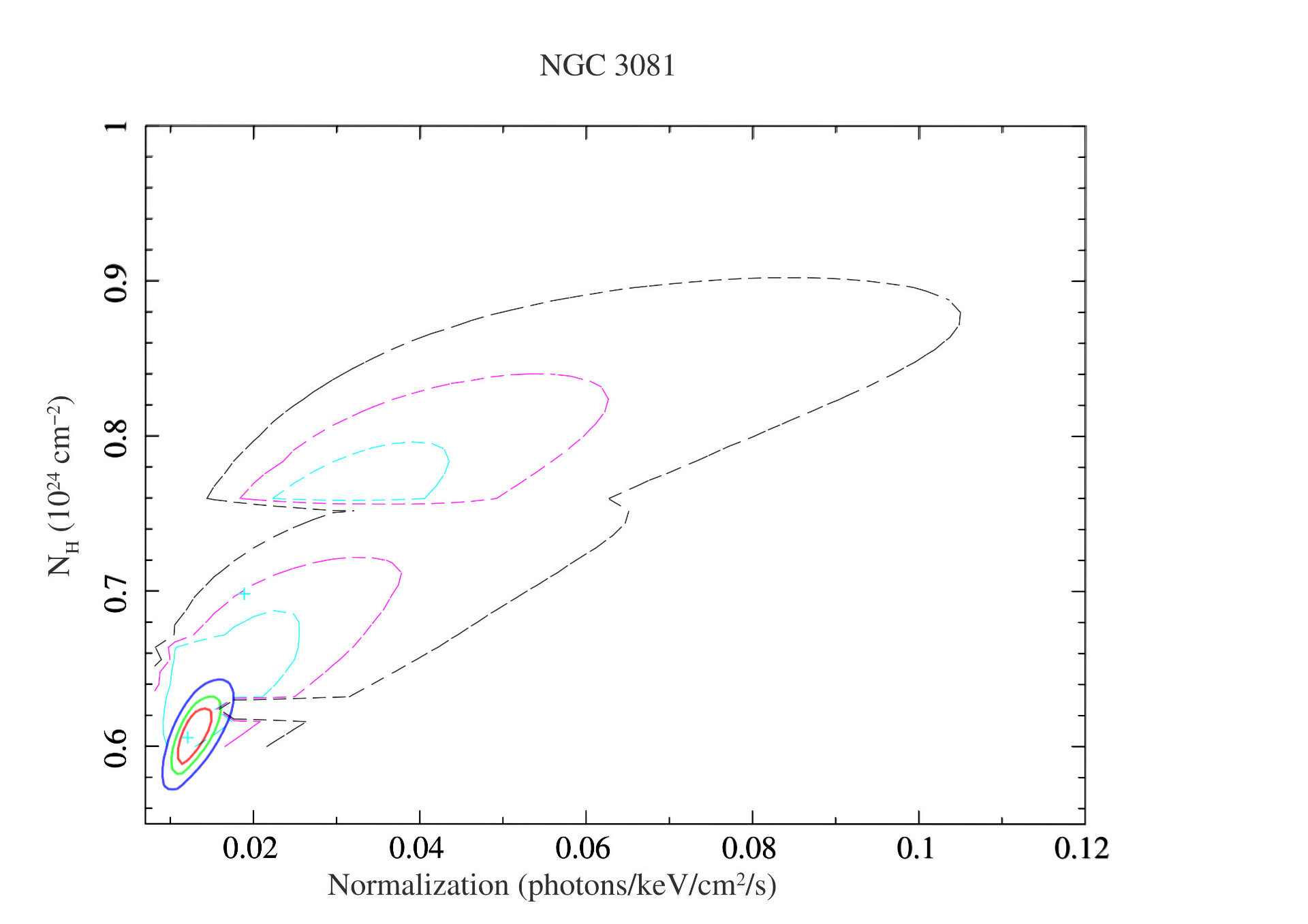}}
\end{minipage}
\caption{\small{Confidence contours for the parameters $\Gamma$-N$_H$ obtained using NGC 3081 spectra in the 0.6--10 keV and 0.6--70 keV energy band (left).
On the right panel we show the confidence contours for the normalization of the continuum and N$_{H,z}$ of the 0.6--10 keV and 0.6--70 keV NGC 3081 spectra superposed. We report with solid lines the \XMM\ + \textit{NuSTAR} contour plots and with dashed lines the \textit{Chandra} ones. The 0.6--70 keV contours are smaller than the \textit{Chandra} ones, due to the larger photon statistic. Moreover, the \textit{Chandra} confidence regions show a double minimum, meaning that there can be two statistically equivalent different combinations for the parameters of interest.}}
\label{fig:ctplot}
\end{figure*}

We investigate possible variability between the \textit{Chandra} data and the \NuSTAR\ + \XMM\ data. While \NuSTAR\ and \XMM\ observations are simultaneous, \cha\ targeted NGC 3081 about one year before. In Figure \ref{fig:ctplot} we show the contour plot between the normalization (indicator of the flux) and the column density (which accounts for the absorption properties). Variability can either be due to intrinsic variation of the emission from the central engine or of the absorbing structure. If there were a difference in both the normalization and in the column density (between the two sets of spectra), we may affirm that it can be due to variation in the geometrical properties of the torus through time (or variation of the accretion efficiency in the case of the normalization). In Figure \ref{fig:ctplot} the superposed \textit{Chandra} and the \XMM\ + \textit{NuSTAR} Normalization-N$_H$ contour plots are shown. The \XMM\ + \textit{NuSTAR} contours plot (solid lines) is much smaller than the \textit{Chandra} one (dashed lines); however, they are consistent, and no variability effects can be attested.
\par We also search for variability effects, for ESO 565-G019, between the \textit{Chandra} and \textit{XMM-Newton}+\textit{NuSTAR} observations through the Normalization-N$_{H,z}$ contour plots. We do not find indications of variability, although we note that the errors on the parameters, due to the limited photon statistics, are large.

\renewcommand{\thefigure}{11}
\begin{figure*}[]
\hspace{10mm}
{\includegraphics[width=1.\textwidth]{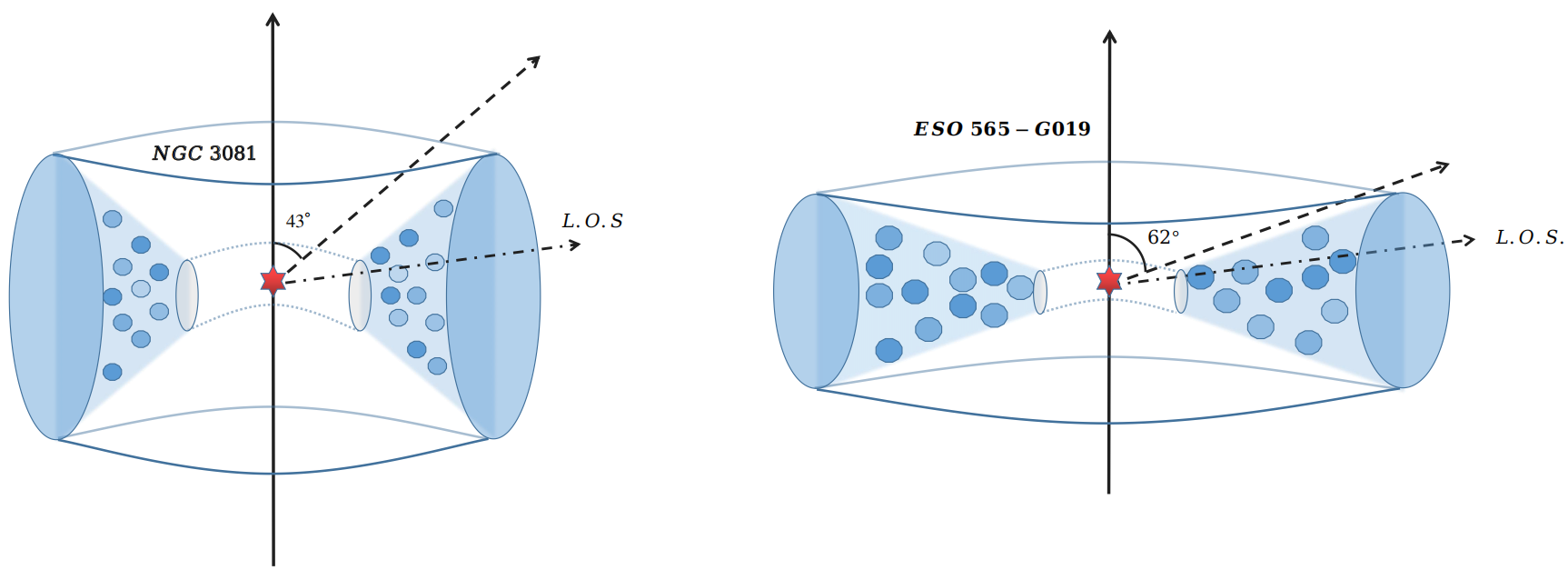}}
\caption{\small{Schematic representation of the two torus configurations found for NGC 3081 (left) and ESO 565-G019 (right). The solid line indicates the torus axis, the dashed line represents the angle corresponding to the torus covering factor and the dash-dotted line is the line-of-sight. The torus clouds, represented as blue circles, are qualitatively color coded with respect to the column density: the darker the color, the higher is the column density.}}
\label{fig:sketch_tori}
\end{figure*}

\subsection{Diffuse emission}

Given the presence of \cha\ observations, it is possible to study the properties of the extended emission in both \ngc\ and ESO 565-G019, thus to establish a better portrait of the soft band spectrum.
\par Following the approach used in \cite{fabbiano2017,jones2020,ma2020},
we obtain the soft (i.e. 0.3--3.0 keV) and hard (i.e. 3.0--7.0 keV) \cha\ images of both sources (see Figures \ref{fig:ngc_doppia} and \ref{fig:eso_doppia}).
We can notice that the diffuse emission extends in the NW-SE and N-S direction, showing an elongated structure on projected scales of about 2 kpc and 3.5 kpc for \ngc\ and ESO 565-G019, respectively.

\renewcommand{\thefigure}{7}
\begin{figure*}[]
\centering
{\includegraphics[width=0.8\textwidth]{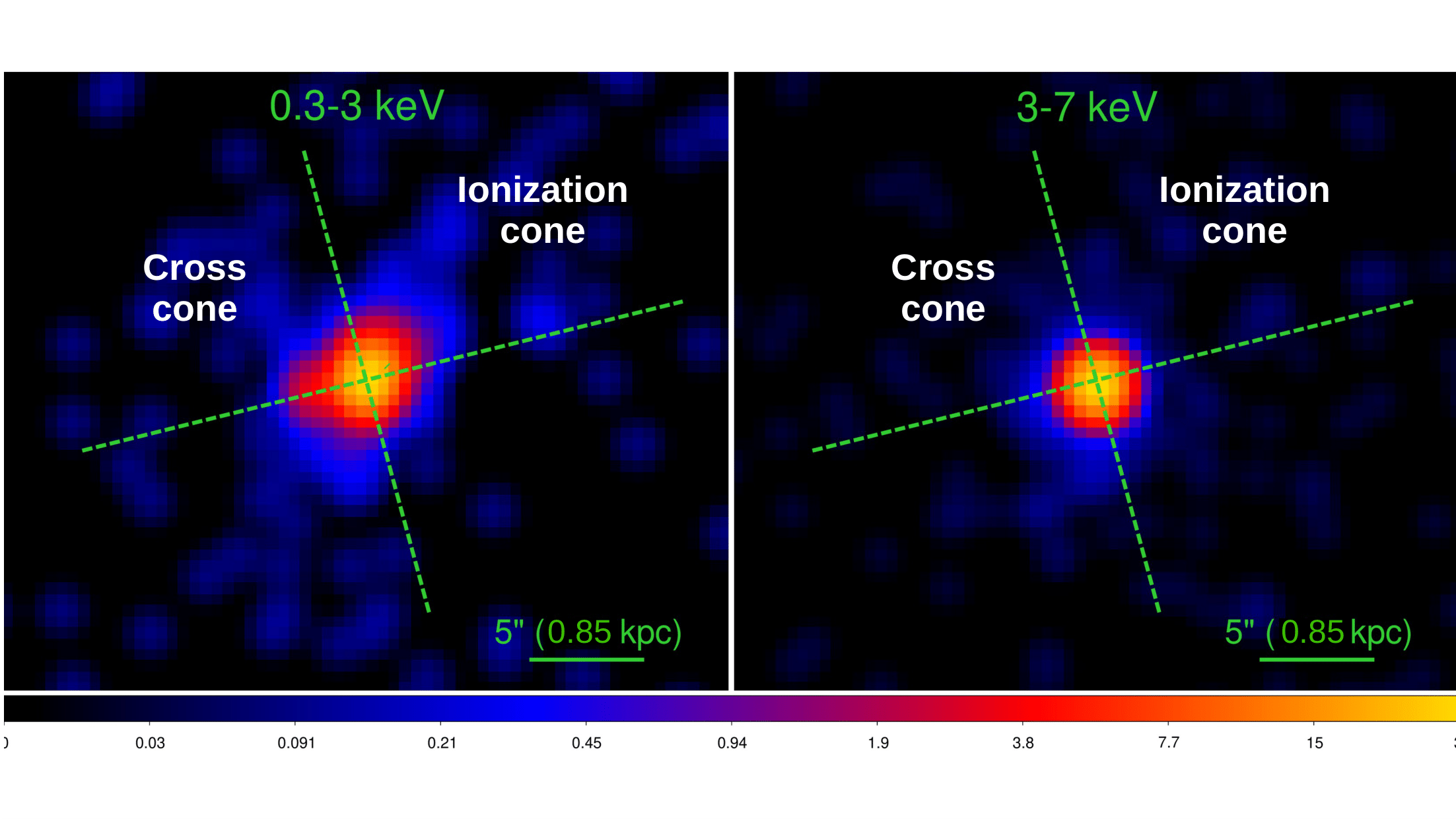}}
\caption{\small{NGC 3081 soft (left) and hard (right) \cha\ images. The green dashed lines indicate the regions where the extended emission is confined. The image is color-coded with the number of counts. Also, the physical scale is reported at the source distance.}}
\label{fig:ngc_doppia}
\end{figure*}

\renewcommand{\thefigure}{8}
\begin{figure*}[]
\centering
{\includegraphics[width=0.8\textwidth]{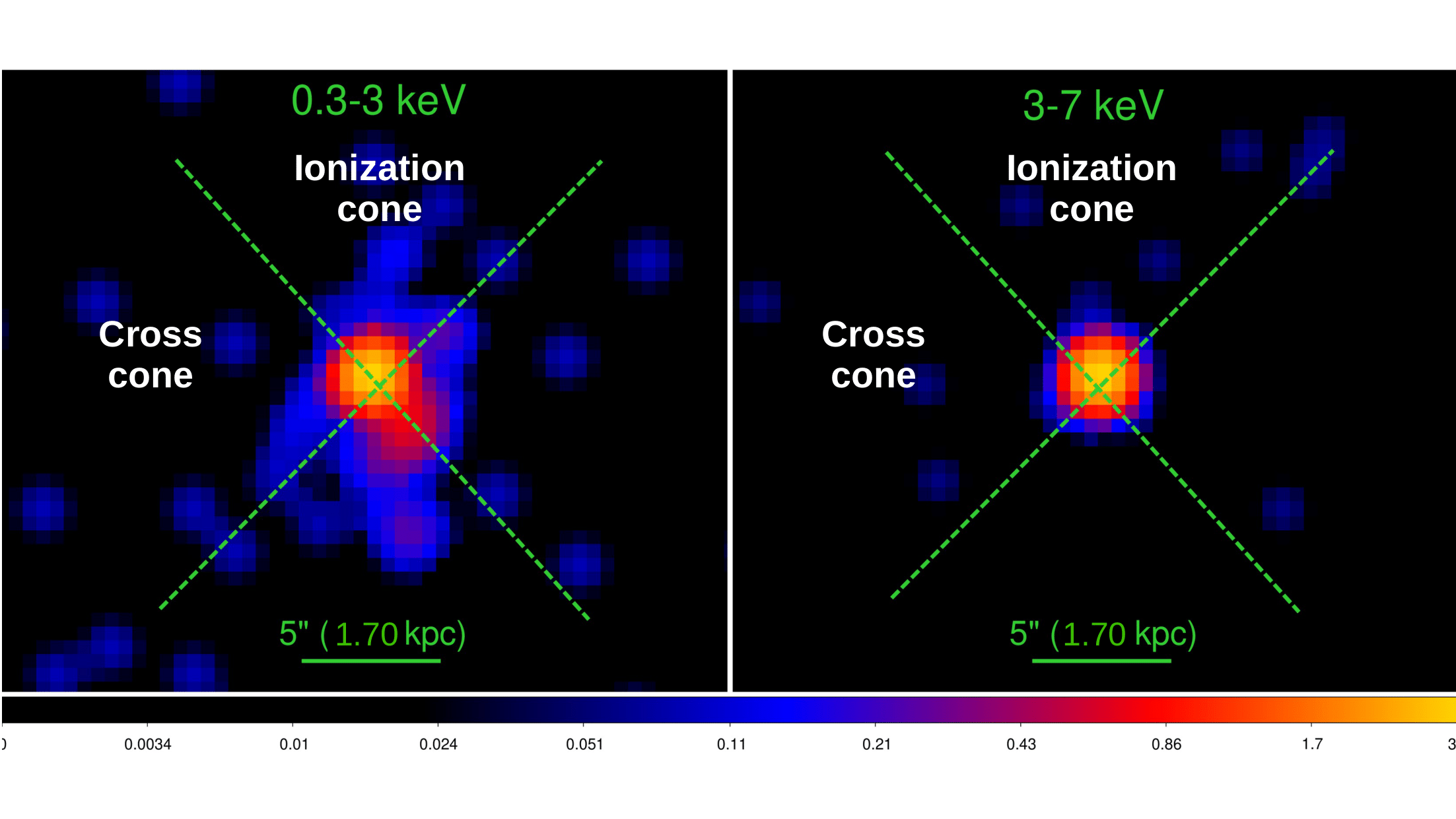}}
\caption{\small{ESO 565-G019 soft (left) and hard (right) \cha\ images. The green dashed lines indicate the regions where the extended emission is confined. The image is color-coded with the number of counts. Also, the physical scale is reported at the source distance.}}
\label{fig:eso_doppia}
\end{figure*}

\par To quantitatively assess the presence (or lack) of extended emission in our sources, we compare the radial distribution of their surface brightness with the one obtained from a simulated PSF in the two energy ranges using the \texttt{ChaRT} and \texttt{MARX 5.5.1} tasks \citep[see e.g.,][for a detailed description of this technique]{fabbiano2017,jones2020,ma2020}. Figure \ref{fig:radial} shows the radial profiles of the emission vs. PSF expectations in the 0.3--3.0 keV and 3.0--7.0 keV energy bands. These profiles have been obtained from an annular region comprising 8 annuli from $\sim 0.7$ to $\sim 8$ arcsec. The 0.3--3.0 keV profiles of \ngc\ and \eso\ show a significant emission above the PSF values up to $\sim 7$ arcsec, whose origin could be related to non-nuclear processes like star formation or diffuse emission on the host-galaxy scales. In Figure \ref{fig:ottico} the \cha\ contours plotted over the optical DSS images are shown. As mentioned in Section \ref{sec:sample}, ESO 565-G019 has a SFR on the order of $\sim$3--4 $M_{\odot}/yr$ \citep[][]{gandhi2013reflection}, thus, its X-ray diffuse emission could be ascribed to a thermal emission on the scales of the host with a possible contribution of star formation processes. Given the 0.5--2 keV spectrum, it is possible to compute the X-ray SFR for these sources following the relation between the 0.5--2 keV luminosity and the SFR \citep[e.g.,][]{ranalli2003SFR}. We find that ESO 565-G019 has a SFR$=4.4_{-0.7}^{+0.8}$ M$_{\odot}/yr$ (the errors correspond to the dispersion of the relation we adopted), consistent with literature.
For NGC 3081 there is little \citep[nuclear SFR $<$0.05 M$_{\odot}$/yr,][]{Esquej2014sfr} to no evidences for nuclear star formation activity \citep[see, e.g.,][]{esparza2018circumnuclear, fuller2019sofia}, therefore the diffuse emission could be produced by galaxy-scale processes (e.g., hot gas in the nuclear region of the galaxy). However, we cannot rule out the possibility that star formation also contributes to some extent. Indeed, from the 0.5--2 keV spectrum, we find a SFR$=1.2_{-0.2}^{+0.2}$ M$_{\odot}/yr$.

In the hard-band profiles, where the AGN contribution is dominant and negligible contamination from non-AGN processes is expected, this extension is much more reduced, especially for ESO 565-G019. We also measure the excess fraction, defined as the ratio between the counts above the PSF and the total counts in the analyzed area. We find an excess fraction of $20 \pm 1.8\%$ for the 0.3--3.0 keV extended emission of NGC 3081 and $18.9 \pm 8.7\%$ for ESO 565-G019. In the 3.0--7.0 keV range, the excess fraction is negative for both \ngc\ ($< -0.01$) and ESO 565-G019 ($< -0.07$), meaning that the emission is consistent with the PSF in the 3.0--7.0 keV energy range.
\par The detection of a diffuse axial emission in the 0.3--7.0 keV interval is in agreement with the aforementioned works \citep[][]{fabbiano2017,jones2020,ma2020}. However, although we detect an excess in the soft data, we find no significant excess in the hard extended emission, which has been found to be 12--22\% and likely due to the existence of
reprocessed emission on scales beyond the torus \citep[see e.g.,][]{ma2020}.

\renewcommand{\thefigure}{9}
\begin{figure*}[]
\begin{minipage}[b]{.5\textwidth}
{\includegraphics[width=1.\textwidth]{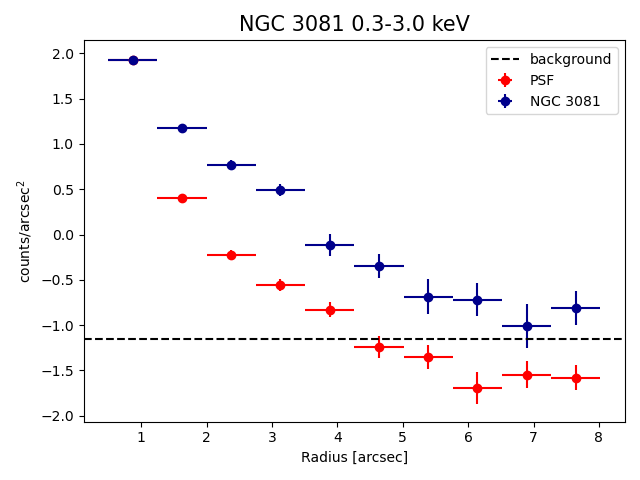}}
\end{minipage}
\begin{minipage}[b]{.5\textwidth}
\centering
{\includegraphics[width=1.\textwidth]{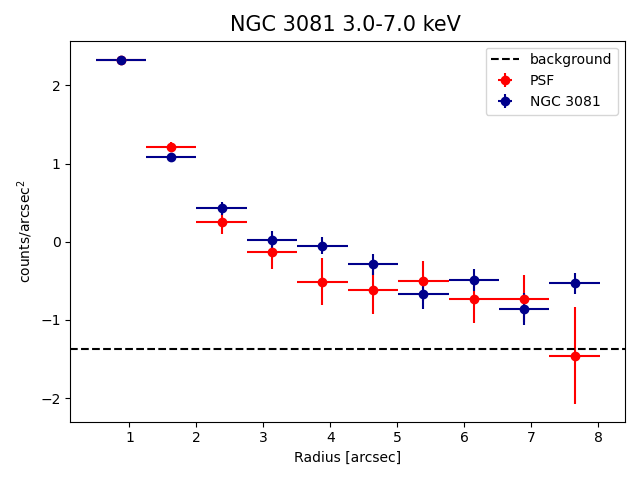}}
\end{minipage}
\begin{minipage}[b]{.5\textwidth}
{\includegraphics[width=1.03\textwidth]{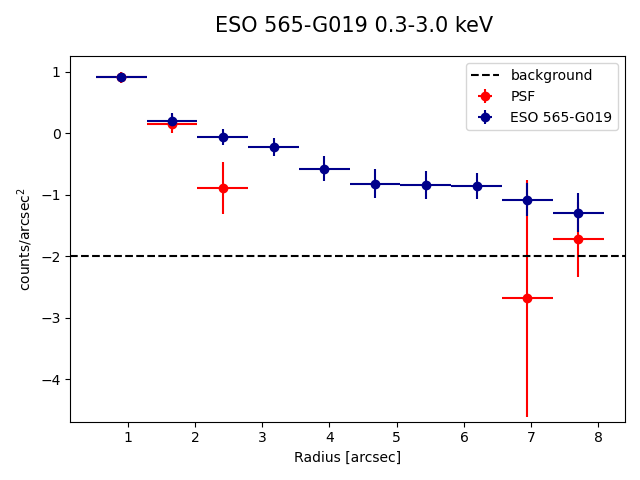}}
\end{minipage}
\hspace{5mm}
\begin{minipage}[b]{.5\textwidth}
\centering
{\includegraphics[width=1.\textwidth]{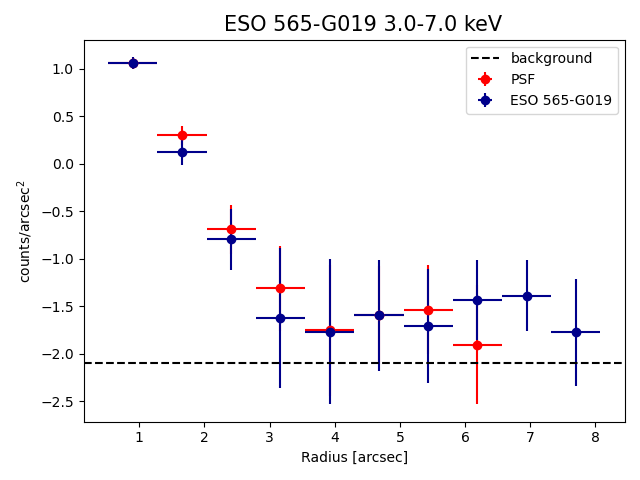}}
\end{minipage}
\caption{\small{Radial profiles in the 0.3--3 keV and 3.0--7.0 keV energy bands for \ngc\ and ESO 565-G019. The blue points are the source counts, whereas the red points represent the counts of the simulated PSF normalized to the first source point. The black dashed line indicates the background level.}}
\label{fig:radial}
\end{figure*}

\renewcommand{\thefigure}{10}
\begin{figure*}[]
\begin{minipage}[b]{.5\textwidth}
{\includegraphics[width=1.\textwidth]{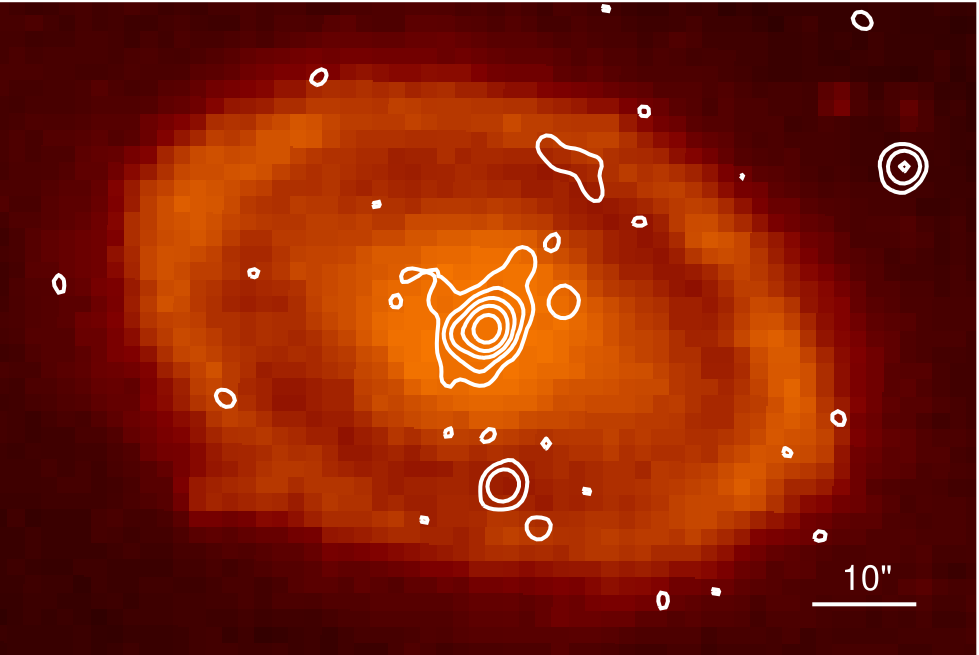}}
\end{minipage}
\hspace{5mm}
\begin{minipage}[b]{.5\textwidth}
\centering
{\includegraphics[width=1.\textwidth]{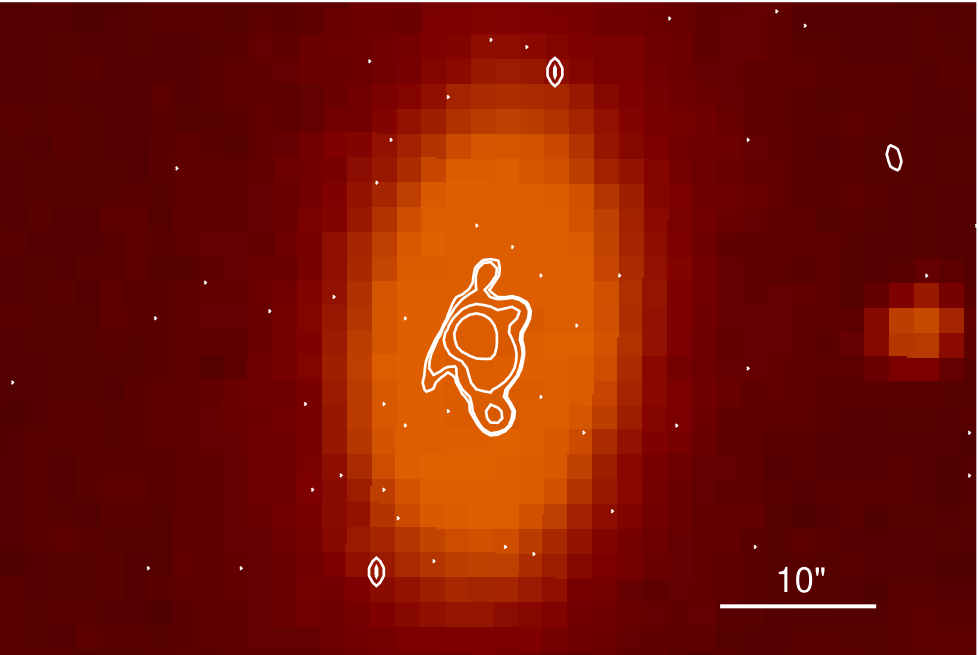}}
\end{minipage}
\caption{\small{Optical DSS images of \ngc\ and ESO 565-G019 in the IIIaJ
band, centered at 4860 \AA. White contours indicates the \cha\ emission in the 0.3-3.0 keV energy range.}}
\label{fig:ottico}
\end{figure*}

\subsection{Comparison with the total sample}

Based on the results of this work along with those reported by Torres-Albà et al. (submitted), eight out of ten sources are incompatible, at 90\% significance, with having the same line-of-sight and average torus column densities. This follows the overall trend observed for the full sample of 57 obscured AGN, in which the large majority ($\sim$91\%) of sources show this discrepancy (see Fig. 4 in Torres-Albà et al., submitted). We link this observational evidence to the presence of a clumpy torus.
Moreover, including the results from the 10 sample analysis, 13 out of 57 candidate CT-AGN (one of which is ESO 565-G019) have both l.o.s and average column density larger than $10^{24}$ cm$^{-2}$. 
With the addition of this source, the percentage of NuSTAR-confirmed CT-AGN in the BAT sample at $z \le 0.05$ is $\sim$8\% (32/417) \footnote{https://science.clemson.edu/ctagn/ctagn/}, still much lower than the predictions. However, with the analysis of the full $\sim$60 source sample, Torres-Albà et al. (submitted) observed a significant decrease of the Compton-thick fraction with redshift. In fact, at $z<0.01$, the fraction is $20.0\pm5.7\%$, which is much closer to the theoretical predictions.

\subsection{Conclusion}
Despite the many studies carried out over the last 20 years, the obscured AGN population is not entirely well characterized and many questions are still unresolved. The study of obscured AGN is relevant for several astrophysical issues concerning galaxy evolution, as well as the CXB content and determination of the accretion history of the Universe \citep[e.g.,][]{alexander2003resolving, gilli2007synthesis, treister2009space}. In fact, the quest for and characterization of Type 2 AGN provides a census of the population of galaxies which are thought to be in the phase of the building up of their mass (probably after a merger) or in the first phases of the nuclear activity \citep[see, e.g.,][]{chen2013correlation, azadi2015primus}. 

\par Combining the capabilities of \cha\ and \XMM\ at E$<10$ keV with those of \NuSTAR\ (3--79 keV), it is possible to properly characterize the spectral parameters (e.g., intrinsic column density and opening angle of the obscuring material) of heavily obscured AGN, allowing to distinguish between the Compton-thin and Compton-thick regime. These combined observations can also break the degeneracy between spectral parameters (e.g., between the photon index and the obscuring material column density) through the use of advanced, physically motivated spectral models (e.g., \MYTorus\ and \borus\ ).
\par In Figure \ref{fig:sketch_tori} a schematic sketch of the best-fit configuration found for the two sources is reported. The torus is represented by several clumps distributed following the toroidal structure. In the figure the following properties are present: the covering factor is represented as the angle of the sky free from the torus by the central source point of view; the differences in the column density of the torus is represented by differences in the colors of the clumps, darker colors mean higher column density. Thus, NGC 3081 has an higher covering factor (lower angle) than ESO 565-G019. The l.o.s. column density of NGC 3081 is lower than the “global average” column density: the observer is looking at the central source through an under-dense region with respect to the torus average column density. ESO 565-G019 is characterized by a Compton-thick column density both on the l.o.s. and in average, and this is represented by over-dense clumps in the whole structure.
\par The conclusions of this work can be summarized as follows:
\\
\\
\textbf{1}. We have verified that the \textit{NuSTAR} data significantly contribute to the determination of the main spectral parameters which characterize obscured AGN at low redshift ($z \le 0.1$). Its contribution mainly consists in the decreasing of the errors of the parameters of interest and in breaking of degeneracy between them, thus allowing a better characterization of the spectral emission properties. Moreover, the use of \textit{XMM-Newton} and \textit{NuSTAR} simultaneous observations allows one to avoid variability effects.
\\
\\
\textbf{2}. The spectra of both sources present a significant emission at energies $< 2-3$ keV which cannot be ascribed solely to the main emission from the nucleus. We have modeled this emission with a thermal component and found that it can be produced by a medium with temperature between $10^6-10^7$ K. However, we do not place constraints on the origin of this emission. This thermal component can be due either to a thermally emitting gas in the nuclear region or to a population of X-ray emitting unresolved sources (e.g., X-ray binaries) or to a combination of the two phenomena. In the soft part of the spectrum of NGC 3081, we find evidences of emission lines, which are typical of obscured sources \citep[e.g.,][]{brinkman2002soft, piconcelli2011x}. 
\\
\\
\textbf{3}. We found that NGC 3081 is best fitted by the \texttt{MYTorus} model in the de-coupled mode (edge-on configuration), while ESO 565-G019 is best fitted by the \texttt{borus02} model. NGC 3081 is Compton-thin along the l.o.s., but with the obscuring material being, on average, Compton-thick. ESO 565-G019 is classified as Compton-thick in both the l.o.s. and average components of the column density.
\\
\\
\textbf{4}. For both sources we were able to compute the torus covering factor through the \texttt{borus02} modeling. NGC 3081 has C$_{TOR}=0.73_{-0.10}^{+0.09}$, suggesting that the torus contributes in a significant way to cover the central emission. Moreover, the ratio between l.o.s. and average column densities is typical of a clumpy scenario. For ESO 565-G019 the lower covering factor C$_{TOR}=0.47_{-0.08}^{+0.18}$, along with the $\Delta$N$_{H}$, suggests a scenario in which the obscuring structure, which is Compton-thick, may be distributed in several individual clouds, responsible for the obscuration. The values are consistent, between the uncertainties, with the average covering factor found by \cite{marchesi2019compton}, C$_{TOR}\simeq 0.6$, for a $\sim$ 30 CT-AGN candidates sample.
\\
\\
\textbf{5}. The main nuclear emission can be divided into the reprocessed component, which is heavily suppressed in Compton-thick AGN, and the component which is scattered (and unabsorbed) by Compton-thin material and can reach the observer at lower energies (i.e. $<5$ keV). For both AGN presented in this work we find that this component represents a low fraction of the main emission, being $<1\%$ and $\sim 1\%$ for NGC 3081 and ESO 565-G019, respectively. These results suggest that the Compton-thin material is a small fraction of the circumnuclear environment.
\\
\\
\textbf{6}. Thanks to the presence of \cha\ data, we were able to investigate the extended emission of NGC 3081 and ESO 565-G019. We found significant diffuse emission in the 0.3--3.0 keV band extending for about 2 (NGC 3081) and 3.5 kpc (ESO 565-G019). However, we were not able to detect any diffuse emission in the 3--7 keV energy range.

%
%
\section*{Acknowledgements}

S.M. acknowledges funding from the the INAF ``Progetti di Ricerca di Rilevante Interesse Nazionale'' (PRIN), Bando 2019 (project: ``Piercing through the clouds: a multiwavelength study of obscured accretion in nearby supermassive black holes''). N.T.A.,  M.A.,  A.P., R.S. and X.Z. acknowledge funding from  NASA  under  contracts  80NSSC19K0531, 80NSSC20K0045 and, 80NSSC20K834.  P.B. acknowledges financial support from the Czech Science Foundation project No. 19-05599Y. M.B. acknowledges support from the YCAA Prize Postdoctoral Fellowship. The scientific results reported in this article are based on observations made by the X-ray observatories \textit{NuSTAR}, XMM-\textit{Newton} and \textit{Chandra} and has made use of the NASA/IPAC Extragalactic Database (NED), which is operated by the Jet Propulsion Laboratory, California Institute of Technology under contract with NASA. We acknowledge the use of the software packages CIAO, XMM-SAS and HEASoft.

\bibliographystyle{aa}
\bibliography{1biblio}

\end{document}